\documentclass[preprint2]{aastex}

\begin{document}
\title{A {\em Chandra}/ACIS Study of 30~Doradus II.  X-ray Point Sources 
in the Massive Star Cluster R136 and Beyond}

\author{Leisa K. Townsley, Patrick S. Broos, Eric D. Feigelson, 
Gordon P. Garmire, Konstantin V. Getman}

\affil{Department of Astronomy \& Astrophysics, 525 Davey
Laboratory, Pennsylvania State University, University Park, PA 
16802} 

\begin{abstract}

We have studied the X-ray point source population of the
30~Doradus star-forming complex in the Large Magellanic Cloud using
high-spatial-resolution X-ray images and spatially-resolved spectra
obtained with the Advanced CCD Imaging Spectrometer (ACIS) aboard the
{\em Chandra X-ray Observatory}.  Here we describe the X-ray sources in
a $17\arcmin \times 17\arcmin$ field centered on R136, the massive star
cluster at the center of the main 30~Dor nebula.  We detect 20 of the 32
Wolf-Rayet stars in the ACIS field.  R136 is resolved at the subarcsecond
level into almost 100 X-ray sources, including many typical O3--O5 stars
as well as a few bright X-ray sources previously reported.  Over two orders
of magnitude of scatter in $L_X$ is seen among R136 O stars, suggesting
that X-ray emission in the most massive stars depends critically on
the details of wind properties and binarity of each system, rather
than reflecting the widely-reported characteristic value $L_X/L_{bol}
\simeq 10^{-7}$.  Such a canonical ratio may exist for single massive
stars in R136, but our data are too shallow to confirm this relationship.
Through this and future X-ray studies of 30~Doradus, the complete life
cycle of a massive stellar cluster can be revealed.

\end{abstract}


\section{INTRODUCTION \label{sec:intro}}

Stars of virtually all masses and stages emit X-rays in their youth,
although the mechanisms for X-ray emission vary with stellar mass.
T-Tauri and protostars have magnetic flares with $L_X$ up to
$10^{31-32}$~ergs~s$^{-1}$ \citep{Feigelson05}.  Individual OB stars
emit X-rays at levels $L_X \sim 10^{-7}$~L$_{bol} \sim
10^{32-34}$~ergs~s$^{-1}$, probably arising from shocks within their
unstable $>1000$~km~s$^{-1}$ winds \citep[e.g.][]{Berghoefer97}.
Colliding winds of binary O and Wolf-Rayet (WR) stars can produce
shocks with $L_X$ up to $10^{35}$~ergs~s$^{-1}$ variable on timescales
of days \citep[e.g.][]{Pollock95}.  For OB stars excavating an H{\sc
II} region within their nascent molecular cloud, diffuse X-rays may be
generated as fast winds shock the surrounding media \citep{Weaver77};
we have recently discovered such parsec-scale diffuse emission with
{\it Chandra} observations of the Galactic high-mass star-forming
regions M17 and the Rosette Nebula \citep{Townsley03}.
 
30~Doradus (30~Dor) is the largest H{\sc II} region in the Local Group,
hosting several young, massive stellar clusters, several well-known
supernova remnants (SNRs), and a vast network of superbubbles created
by current and past generations of massive stars and their supernovae.
At the center of 30~Dor is the massive compact stellar cluster R136
\citep[][called ``RMC~136'' in SIMBAD]{Feast60}, the richest resolved
OB association with dozens of 1--2~Myr-old $> 50$~M$_\odot$ O and WR
stars \citep{Massey98}, including several examples of the
recently-defined earliest spectral type, O2 \citep{Walborn02}.
Including stars down to $0.1$~M$_{\odot}$, R136 has a mass of $\sim 6
\times 10^4$~M$_{\odot}$ \citep{Brandl05}; there are $>$3500 stars
within its 10-pc diameter \citep{Massey98}.

30~Dor contains other stellar clusters 1--10 million years old, the
$\sim 20$~Myr old stellar cluster Hodge~301 \citep{Grebel00}, and a new
generation of deeply embedded high-mass stars just now forming
\citep[][and references therein]{Walborn02b}, making it a prime example
of sequential and perhaps triggered star formation.  Recent radio
observations have revealed water \citep{Loon01} and OH \citep{Brogan04}
masers in 30~Dor, also signifying ongoing star formation.  The presence
of these young objects as well as a large population of
widely-distributed WR stars \citep{Moffat87} and evolved supergiants
$\sim 25$~Myr old \citep{Walborn97} shows that 30~Dor is the product of
multiple epochs of star formation.  The new generation of embedded
stars currently forming may be the result of triggered collapse from
the effects of R136 \citep{Brandner01}.

\begin{figure}
\centering
  \includegraphics[width=1.0\textwidth]{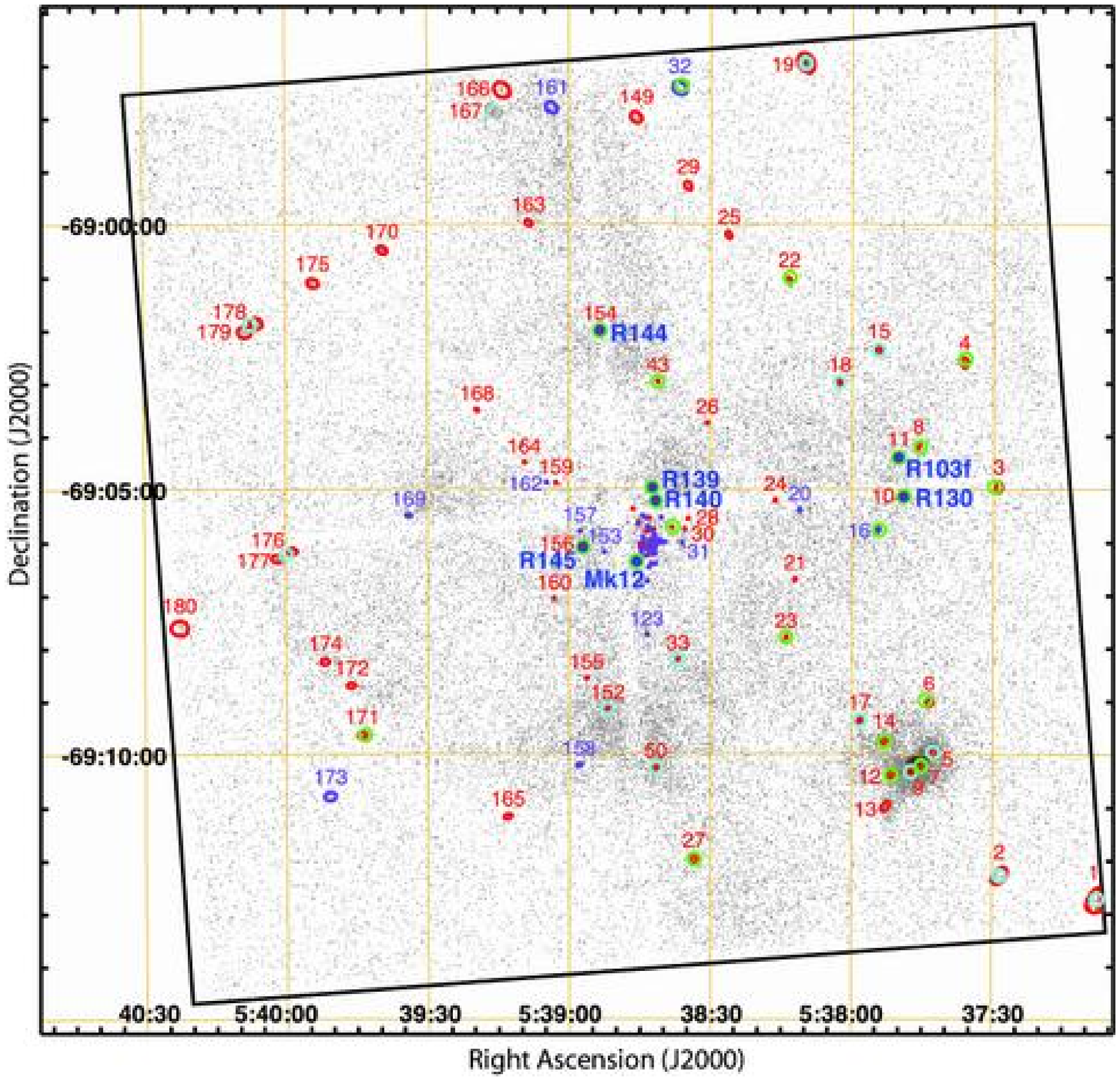}
\caption{A binned image showing the {\em Chandra} observation of 30~Dor
on the $17\arcmin \times 17\arcmin$ACIS-I array.  The aimpoint was on
R136.  Data were binned by four sky pixels ($2\arcsec \times
2\arcsec$); events with energies between 0.5~keV and 7~keV were
included.  {\em Chandra} point sources are outlined with red polygons
(primary sources from Table~\ref{tbl:sourcelist}) and purple polygons
(tentative sources from Table~\ref{tbl:tentative}) and are labeled with
their sequence numbers from those tables.  These polygons define the
event extraction regions for each source and were generated by {\it
ACIS Extract} using the point spread function (PSF) appropriate for
each source location.  Their shapes reflect the shape of the PSF at
each source location.  Some counterparts from other studies are shown
with colored circles: 2MASS in green, previously known X-ray sources in
cyan, WR stars and a few other notable sources in blue (see
\S\ref{sec:ptsrcs}).  Counterparts in R136 (field center) have been
omitted for clarity.
\label{fig:bin4data}} 
\end{figure}

We observed 30~Dor with {\em Chandra's} Advanced CCD Imaging
Spectrometer (ACIS) camera for $\sim 21$~ks in 1999 September; see
Figure~\ref{fig:bin4data}.  R136 was positioned at the aimpoint of the
$17\arcmin \times 17\arcmin$ ACIS Imaging Array (ACIS-I).  The reader
is referred to the accompanying paper \citep[][henceforth Paper
I]{Townsley05} for a review of previous X-ray studies of the 30~Dor
complex, a description of our {\em Chandra} observations and data
analysis, and a detailed treatment of the superbubbles, SNRs, and other
diffuse structures in the field.  Here we consider the point sources
located in the stellar clusters and distributed across the field.  Our
methods for detecting and extracting sources are described in
\S\ref{sec:srcdetect}.  The X-ray properties of these sources and
counterparts identified from other studies are tabulated in
\S\ref{sec:ptsrcs}.  All luminosities are calculated assuming a
distance of 50 kpc.

Many of the 180 ACIS-I point sources identified in this observation are
early-type stars or stellar systems associated with the very dense,
massive stellar cluster R136 (seen as a large concentration of sources
at the center of Figure~\ref{fig:bin4data}).  Our efforts to resolve
and identify sources in R136 are detailed in \S\ref{sec:R136}.  Several
point sources are spatially associated with the N157B SNR; some may be
clumps in the diffuse emission, but the cospatial OB association LH~99
may account for true point sources in this region.  Several dozen faint
sources are scattered across the field, notably the well-known WR stars
R130, R134, R139, R140a, R140b, R144, and R145.  Most X-ray sources
without visual or infrared (IR) counterparts are likely background
active galactic nuclei (AGN), but some appear to have unusually hard
spectra and may be low-luminosity X-ray binaries associated with the
30~Dor complex.  These distributed sources are described in
\S\ref{sec:otherpoints}.

\section{SOURCE DETECTION AND CHARACTERIZATION METHODS
\label{sec:srcdetect}}

Starting with the three Level 1 event lists that make up this unusual
dataset (see Table 1 in Paper I), we performed several custom data
processing steps to filter and improve the individual event lists, as
described in Paper I.  Source detection was then performed on the
merged data using {\it wavdetect} \citep{Freeman02} with a threshold of
$1 \times 10^{-5}$ due to the presence of diffuse emission
\citep{Townsley03}.  Source searching was performed in three bands
(0.5--2 keV, 2--8 keV, and 0.5--8 keV) using images binned at 1, 2, and
4 ``sky pixels'' ($0\arcsec.5$, $1\arcsec$, and $2\arcsec$
respectively) and the resulting 9 sourcelists merged.  Considering only
sources on the ACIS-I array, this series of {\it wavdetect} runs
resulted in a list of 167 potential point sources.  Inspecting the data
showed that several of these potential sources were actually clumps in
the diffuse emission and that {\em wavdetect} had missed several likely
sources in the core of R136.

Due to the high density of point sources in the R136 core, the ACIS
data suffer from source confusion in this region.  Since this cluster
was placed at the aimpoint of the ACIS-I array, R136 sources have
sharply-peaked, sub-arcsecond Point Spread Functions (PSFs), so our
knowledge of the X-ray emission of those sources might benefit from
super-resolution techniques.  Pursuant to this idea, we applied ACIS
sub-pixel resolution software to the photon events \citep{Mori01},
formed several high-resolution images of the R136 region from those
events, reconstructed those images using a simple maximum likelihood
algorithm (described below), and used the reconstructions to identify
additional X-ray point sources in R136.  Similar reconstructions have
been performed on ACIS observations of SN1987A \citep[e.g.][]{Park05}
and gravitational lenses \citep[e.g.][]{Chartas04} and revealed
subarcsecond spatial features that were confirmed in other wavebands
and proved essential for understanding the X-ray emission of these
targets.

Only the ACIS-I array was considered in our point source searching and
characterization; the S3 and S4 chips in the ACIS Spectroscopy Array
also show emission consistent with point-like sources (see Figure 2 in
Paper I), but the large PSF at these locations ($>20\arcmin$ off-axis)
and confusion with the extensive diffuse emission there makes
quantitative analysis of these sources difficult.  As described in
Paper I, other {\em Chandra} and {\em XMM-Newton} observations are
better-suited to the analysis of these sources.

\subsection{Image Reconstruction and Source Identification
\label{sec:srcid}}

To elucidate the source population in R136 and in the subgroup R140
\citep{Moffat87}, we performed maximum likelihood image reconstructions
on these fields using the IDL routine {\it max\_liklihood.pro} in the
IDL Astronomy User's Library maintained by Wayne
Landsman\footnote{\url{http://idlastro.gsfc.nasa.gov/homepage.html}}.
These reconstructions are built into PSB's {\it ACIS Extract}
software\footnote{The {\it ACIS Extract} code is available at
\url{http://www.astro.psu.edu/xray/docs/TARA/ae\_users\_guide.html}.}
that performs source extraction and automated spectral fitting
\citep{Broos02}.  The user can perform maximum likelihood image
reconstruction for every extracted source if desired; the code builds
an image of the source ``neighborhood'' (a region $\sim 50\arcsec
\times 50\arcsec$ in size for on-axis sources) and uses that source's
PSF at an energy specified by the user (1.5~keV in this case) obtained
from the CALDB's PSF library in the reconstruction.  Since the true
{\it Chandra} PSF varies substantially with off-axis angle, we chose to
reconstruct the neighborhoods around two sources in R136 separated by
$\sim 30\arcsec$ in order to cover the entire region suffering from
source confusion using appropriate PSFs.  Our goal was to supplement
the candidate list of point sources in very crowded parts of the field,
where both {\em wavdetect} and visual examination fail to provide a
good census of sources. 

While the R140 reconstruction did not reveal any new point sources in
that cluster, the R136 reconstruction separated the unresolved emission
in this region into many peaks, hinting that many more X-ray point
sources were present in this region than the $\sim 10$ that {\it
wavdetect} revealed (see \S\ref{sec:ptsrcs}).  We quantified peaks in
the reconstructed image using the IDL Astronomy User's Library routine
{\it find.pro}, an IDL version of the {\it find} routine in Peter
Stetson's DAOPHOT package \citep{Stetson87}.  This routine gives the
centroid position of each peak and a measure of its shape and
brightness.  Peaks that were clearly associated with single events in
the data were filtered out using these shape and brightness criteria.
Then using {\it SAOImage
DS9}\footnote{\url{http://hea-www.harvard.edu/RD/ds9/}}, we overlaid
apertures representing the 90\% on-axis PSF contour onto an image of
the ACIS data, placed at the locations of reconstruction peaks.  Only
those peaks that contained at least two ACIS events within the 90\%
aperture were kept for further consideration.  This list of potential
sources from maximum likelihood reconstruction was merged with that
from {\it wavdetect}, yielding 288 candidate sources on the ACIS-I
array.

We then performed a preliminary event extraction for these candidate
sources using {\it ACIS Extract}, which includes a calculation of the
probability that the events contained in a given extraction region are
due solely to Poisson variations in the local background (a quantity
that we will later call $P_B$ in
Tables~\ref{tbl:sourcelist} and \ref{tbl:tentative}).  Note that we did
not allow reconstruction peaks that were generated from single isolated
X-ray events to be considered X-ray point sources, regardless of their
statistical likelihood, due to the presence of diffuse emission in the
region.  This probability provides an objective estimate of source validity;
at this stage of the analysis, we required that each candidate source have no
more than a 1\% likelihood of being a background fluctuation in order
to be considered valid.  This resulted in a final list of 180
acceptable potential sources\footnote{Note that the validity of an
individual point source $S$ often cannot be judged without
consideration of the full catalog of sources.  The validity of $S$
depends on the local background estimate for $S$ which must obviously
exclude regions thought to contain other point sources.  Thus members
of the catalog must be chosen before their final source validity is
calculated, and source validity estimates are an obvious way to judge
membership in the catalog.  We have not attempted to define a source
identification and local background estimation process that results in
a completely unique solution.  The initial set of 288 potential sources
was extracted and then cut at the threshold of 1\% likelihood of being
a background fluctuation.  This trimmed list of 180 potential sources
was then re-extracted, resulting in different local background
estimates and different source validity metrics $P_B$.  Thus some $P_B$
values in Table~\ref{tbl:tentative} are $>$ 1\% ($\log P_B > -2$).},
ranging in brightness from 1.3 to 6187 net counts.  Since this source
validity test relies on local background estimates, extremely faint
sources can be reliably detected in regions of very low background;
such conditions exist in the R136 core.  Given the extraordinarily high
source density in R136, though, even the reconstructed sources may
actually be source blends.  

Local background estimates, hence $P_B$ used in an absolute sense, are
subject to substantial uncertainty in our sparse datasets.  In 30~Dor
this is complicated by the extensive, highly-structured diffuse
emission in the field;  to get a reasonable number ($\sim 100$) of
background counts one must travel far ($\sim 15\arcsec$ for our on-axis
sources) from the source, making the background estimate not as
``local'' as one might wish.  We use $P_B$ as a way to make reasonable
distinctions between our primary sourcelist
(Table~\ref{tbl:sourcelist}, in which we have substantial confidence),
our tentative sourcelist (Table~\ref{tbl:tentative}, containing sources
still quite likely to be real), and failed candidate sources.  

To test the stability of the primary sourcelist definition criterion
($P_B < 0.003$) to changes in the local background, we deleted the 33
sources in Table~\ref{tbl:tentative} with $P_B > 0.01$ (leaving 30
tentative sources) and re-extracted the remaining 147 sources,
re-calculating $P_B$ for each source.  Most of the deleted sources were
in R136; when they were removed from the input sourcelist, {\it ACIS
Extract} treated their photons as background.  We wanted to test
whether this added background substantially altered the detection
significance of the remaining R136 sources, as a way to test the
fidelity of using image reconstruction to nominate sources for the {\it
ACIS Extract} process.  After this second extraction, 32 of the 147
sources had $P_B > 0.003$ and would be considered tentative; of those
32, 15 had $P_B > 0.01$.  A few sources with $P_B$ near the cut-off
value of 0.003 changed tables:  four primary sources (ACIS \#33, 94,
106, and 116) moved to the tentative table and two tentative sources
(ACIS \#97 and 123) moved to the primary table.  All other primary
sources continued to have $P_B < 0.003$; this demonstrates that our
method of selecting high-quality point sources using {\it ACIS Extract}
is stable to changes in the tentative sourcelist and local background.
Thus we retained the original 180 potential X-ray point sources for
further analysis, divided into primary and tentative sourcelists based
on $P_B$ from the first extraction.  Obviously, users of these data
should view tentative sources with appropriate caution based on the
value of $P_B$ for each source.

\subsection{Source Extraction}

With the caveats of the last section in mind, we extracted a final set
of source characteristics for our trimmed list of 180 sources.
Positions for the sources derived from reconstructions were unaltered;
positions for {\it wavdetect} sources were recomputed during the
preliminary extractions.  Isolated sources were assigned extraction
regions corresponding to the 90\% contours of their PSFs.   Crowded
sources were assigned extraction regions corresponding to smaller (down
to 40\%) PSF fractions to ensure that virtually no photons were
included in more than one source.  Note however that properties of
crowded sources may still be affected by the PSF wings of their
neighbors.  Source spectra, ARFs, and RMFs were constructed by {\it
ACIS Extract}, which uses standard CIAO tools.

Spatial masking was applied to the event data and exposure map in order
to eliminate most point source light prior to extraction of local
background spectra for each extracted source.  Such masking and
background computations necessarily involve a tradeoff between the
competing goals of eliminating all the light from the point sources and
making the background spectra as ``local'' as possible.  We chose a
relatively sophisticated iterative approach which seeks to mask all
regions where the expected surface brightness from the point sources is
larger than one half the observed background level.  An expected
surface brightness image is computed using the exposure map and source
flux estimates from preliminary extractions.  A smooth background image
is iteratively computed by smoothing the masked event data and exposure
map.  The resulting mask always covers all the source extraction
regions, and extends beyond them for bright sources.

Local background spectra for each source were extracted from the masked
event data and exposure map by searching for a circular aperture
centered on the source that contained at least 100 counts and included
an unmasked area at least 4 times that of the source region.  The
background spectra were scaled appropriately using the masked exposure
map pixels found in the background and source regions.  Typically the
scaled full-band background level in on-axis spectra is $<1$ count.

\subsection{Source Reliability and Limiting Sensitivity
\label{sec:sensitivity}}

It is difficult to evaluate rigorously the reliability of this
sourcelist.  Since this region may contain diffuse emission from the
surrounding superbubbles or from R136 itself, some of the reconstructed
sources may be spurious or may be source blends.  This is undoubtedly
true for the X-ray source coincident with the cluster core R136a.  We
have assessed the validity of these sources by using {\it ACIS Extract}
to calculate the probability that any of them is a background
fluctuation, as described above.  The 117 sources with less than a
0.3\% likelihood of being a background fluctuation constitute our
primary sourcelist (Table~\ref{tbl:sourcelist}), while the 63 sources
that have more than a 0.3\% chance of being spurious background
fluctuations but otherwise appear legitimate are listed as tentative
sources in Table~\ref{tbl:tentative}.  These tables are described in
detail in \S\ref{sec:ptsrcs}.  We note that 33 of the 63 tentative
sources have $P_B > 1$\%; those sources in particular should be treated
with caution.  All images of the ACIS data shown in this paper show
primary source extraction regions outlined in red, while tentative
sources are outlined in purple.

Given the highly structured diffuse emission present in this field, a
few of the sources remaining in our list may be knots in this diffuse
emission rather than truly point-like sources.  This is much more
likely to happen far from the center of the field, where PSFs are
large.  In a longer observation, sources could be further tested for
validity by considering their spectra.  {\it ACIS Extract} performs a
2-sided Kolmogorov-Smirnov test that could be used to make sure that
the spectra extracted from a source and its local background are not
identical.  For this dataset, however, most sources are too faint to
use this test.

We estimate the limiting sensitivity of the source list in the on-axis
region around R136 using Koji Mukai's
PIMMS\footnote{\url{http://xte.gsfc.nasa.gov/Tools/w3pimms.html}}
tool.  To avoid the X-ray Eddington bias defined by \citet{Wang04}, we
consider a full-band countrate limit of 5 counts in 21870 sec (the
total for all observations on CCD3).  Assuming $N_H = 5 \times
10^{21}$~cm$^{-2}$ (a rough average of the spectral fits in
Table~\ref{tbl:apec-spec}), the observed X-ray flux for a range of
thermal plasmas with $kT =$ 1--3~keV is $F_X = 1$--$2 \times
10^{-15}$~ergs~s$^{-1}$~cm$^{-2}$.  This gives an intrinsic
(absorption-corrected) source flux of $F_{X,corr} \sim 3.4 \times
10^{-15}$~ergs~s$^{-1}$~cm$^{-2}$, or a limiting full-band luminosity
of $\sim 1 \times 10^{33}$~ergs~s$^{-1}$ in the LMC.  The observation
is thus not sufficiently sensitive to detect individual low-mass
pre-main sequence stars, as their X-ray luminosities rarely exceed $1
\times 10^{32}$~ergs~s$^{-1}$ \citep{Feigelson05}.  From a {\em ROSAT}
survey of early-type stars \citep{Berghoefer97}, this roughly limits
our detection of individual stars on-axis to $\sim$O5 and earlier.

\section{RESULTING X-RAY POINT SOURCES}
\label{sec:ptsrcs}

As described above, we detect 180 pointlike sources in this ACIS
observation; their extraction regions are outlined in red (primary
sources) and purple (tentative sources) in Figure~\ref{fig:bin4data}
and the ACIS sequence number from Tables~\ref{tbl:sourcelist} and
\ref{tbl:tentative} is shown for sources outside of the crowded R136
region.  This binned image of the ACIS-I array also shows some of the
counterparts to our {\em Chandra} sources:  green circles indicate
2MASS sources and cyan circles show some of the X-ray sources known
from previous studies.  Counterparts in the dense central cluster have
been omitted for clarity; a full list of counterparts is given in
Table~\ref{tbl:counterparts}, which is described in detail below.  The
blue labeled counterparts are a mix of interesting sources:  R103f is a
foreground M star, Mk~12 is a B supergiant, and the others are WR
stars.

\citet{PPL02} also analyzed part of this {\it Chandra} observation of
30~Dor.  We confirm their detections of X-ray sources except for their
sources CX13 and CX16, which did not pass our criteria to be legitimate
sources.  They note that their CX1 is a blend of the R136a core and
that the source they find associated with R140 (CX10) is extended.  We
confirm these statements and in fact find that all nine of the sources
that they identify in R136 have faint neighbors that fall within the
90\% PSF contour of those sources.  Our spectral fitting is in rough
agreement with their results for the sources that we have in common.

\subsection{Source Properties}

Basic properties of the 180 pointlike sources are given in
Table~\ref{tbl:sourcelist} (for the 117 primary sources) and Table~\ref{tbl:tentative} (for the 63 tentative sources).  The format of
these tables closely follows that of \citet{Getman05} Tables 2 and 4;
details on how certain columns were derived are given there.  Column 1
is a running sequence number; Column 2 gives the formal source name
based on its J2000 sexagesimal coordinates.  The J2000 coordinates in
decimal degrees are given in Columns 3 and 4.  

The formal $1\sigma$ radial positional uncertainty based on source
counts is given in Column 5.  It is calculated by {\it ACIS Extract}
using the following steps:  (1) assume the parent distribution of the
extracted counts is the  observatory PSF within the source's extraction
region; (2) compute the standard deviations of that truncated PSF along
the sky coordinate axes; (3) divide those standard deviations by the
square root of the number of counts extracted; (4) sum the resulting
single-axis positional uncertainties in quadrature.  The resulting
radial positional uncertainties, converted to units of arcseconds, is
reported in Column 5.

Positional uncertainties can also be estimated from the Chandra Orion
Ultradeep Project (COUP), which provides a large collection of ~1400
sources associated with young stars across the full ACIS-I field of
view \citep[e.g.][]{Getman05}.  Source positions for COUP are based on
the same {\it ACIS Extract} algorithms used here and the COUP
observations were first registered to the 2MASS astrometric reference
frame.  We examined $\sim 350$ COUP sources with astrometric
near-IR counterparts and fewer than 100 net counts for
ACIS$-$star positional offsets.  For sources lying within $5\arcmin$ of
the aimpoint where the {\em Chandra} PSF is excellent, we find average
offsets of $0\arcsec.2$--$0\arcsec.3$ for sources with $3<NetCts<100$.
Some of this is likely due to uncertainties in the IR positions
which are estimated to be $0\arcsec.1$--$0\arcsec.2$.  For sources
lying $5\arcmin$--$7\arcmin$ off-axis, average offsets are $1\arcsec.0$
for $3<NetCts<10$ and $0\arcsec.6$ for $11<NetCts<100$.  For sources
$7\arcmin$--$10\arcmin$ off-axis, average offsets are $2\arcsec$ for
$3<NetCts<20$ and $0\arcsec.8$ for $21<NetCts<100$.  These COUP
positional uncertainties should apply to 30~Dor sources as well and
give a sense of the systematic errors that apply in addition to the
formal position errors given in Tables~\ref{tbl:sourcelist} and
\ref{tbl:tentative}.

Since the {\em Chandra} PSF is a strong function of radial distance
from the aimpoint, the off-axis angle of each source is given in Column
6.  Columns 7--11 give the source photometry, listing each source's net
counts in the full (0.5--8~keV) and hard (2--8~keV) spectral bands,
$1\sigma$ errors on those net counts based on \citet[][Equation
7]{Gehrels86}, the estimated number of background counts in the source
extraction region, and the fraction of the PSF used to extract each
source.  Sources are extracted with a nominal 90\% PSF fraction unless
they are crowded; thus a value substantially less than 90\% in Column
11 indicates that the source was crowded and that its properties may be
influenced by nearby sources.

Column 12 gives the source's photometric significance, defined as the
net counts value (Column 7) divided by the error on that value (Column
8).  Column 13 gives the probability that the source is a background
fluctuation rather than a true point source.  This represents our
primary test of source validity.  As described in \S\ref{sec:srcid},
this quantity differentiates the main source catalog in
Table~\ref{tbl:sourcelist} (117 sources with final $P_B \leq 0.003$)
from the list of tentative sources in Table~\ref{tbl:tentative} (63
sources with final $P_B > 0.003$).

Observational anomalies are noted by a set of flags in Column 14.  For
this field, the only anomalies noted were sources falling in the gap
between the CCD chips or on the edges of the field; both effects cause
sources to have reduced exposure time and edge sources have poor
position estimates due to truncated PSFs.  Column 15 is a
characterization of each source's likelihood of displaying variability
in this observation.  It is based on a calculation of the
Kolmogorov-Smirnov statistic and was first defined in
\citet{Feigelson04}.  This quantity is omitted for sources falling in
chip gaps or on field edges because satellite dithering can cause such
sources to have erroneously variable lightcurves; the test is not
performed for very faint sources.  Column 16 gives the source's
``effective'' exposure time, namely the length of time that the source
would have to be observed at the {\em Chandra} aimpoint in order to
collect the same number of counts as were actually obtained in the
observation performed.

Finally, an estimate of the source's background-corrected median energy
in the full spectral band completes the table.  To compute this
statistic we first construct a cumulative energy distribution
normalized to unity by integrating the background-corrected spectrum
and dividing by net counts.  We then search for the lowest and highest
energies at which the cumulative distribution crosses 50\% -- these
energies may be different because the background-corrected cumulative
energy distribution may not be monotonic (faint sources may have channels
with negative values because that channel contained no source counts but
it did contain background counts).  The average of these two
energies is reported as the median of the background-corrected
spectrum.

\subsection{X-ray Spectroscopy}

Spectra, backgrounds, ARFs, and RMFs for all point sources were
constructed by {\it ACIS Extract} and used for both automated and
interactive spectral fitting using Keith Arnaud's {\it XSPEC} fitting
package \citep{Arnaud96} and our {\it XSPEC} scripts.  Only about a
quarter of these sources are bright enough to yield sufficient counts
for spectral fitting in this short observation.  Sources with
photometry significance $>2$ (49 sources in total, all from the primary
sourcelist in Table~\ref{tbl:sourcelist}) were fit using CSTAT
(ungrouped spectra) in {\it XSPEC12} with a single {\em apec} thermal
plasma model \citep{Smith01} with a single solar-abundance absorption
component \citep[{\em wabs} model,][]{Morrison83}.  If the thermal
plasma fit was unacceptable because it gave non-physical parameters or
failed to converge and the source was not spatially associated with
R136, a power law model was used instead, on the assumption that the
source could be an X-ray binary or a background AGN.  Sources in R136
were only fit with the thermal plasma model because we assume that
these are normal stars, so the thermal plasma model is more
appropriate. 

We chose to use the simple {\em wabs} absorption model to facilitate
future comparison with Galactic star-forming regions.  The assumption
of solar abundance for the absorbing material may be significantly in
error, as most of the absorption is thought to be local to the 30~Dor
nebula \citep{Norci95}.  \citet{Mignani05} note, as part of their
spectral fitting to ACIS data of PSR~J0537$-$6910 in N157B, that
reducing the abundance of the absorbing material to 0.4$Z_{\odot}$
doubles the resultant $N_H$.  We caution the reader that this
systematic underestimate of $N_H$ due to our assumption of solar
abundances for the intervening material pertains to all of our fitting
results.  
  
Tables~\ref{tbl:apec-spec} and \ref{tbl:powerlaw-spec} summarize our
{\it XSPEC} fit results for those sources with photometry significance
$>2$.  Fits with high absorption ($\log N_H >22.5$) omit the
soft-band luminosity $L_S$ and the absorption-corrected full-band
luminosity $L_{t,c}$ because these quantities cannot be determined
reliably.

Fits to the fainter sources in this group should be treated with
appropriate skepticism.  For example, source \#164 in
Table~\ref{tbl:apec-spec} (CXOU~J053909.46$-$690429.8) has an
acceptable fit, but the soft thermal plasma combined with the large
absorbing column yields a very large intrinsic luminosity estimate,
suggesting that this 10-count source is the third brightest pointlike
object in the field, bested only by the pulsar and cometary nebula
source in N157B.  Attempts to fit the spectrum with a lower absorbing
column and different spectral model (including a power law) were
unsuccessful.  The nature of CXOU~J053909.46$-$690429.8 is unknown so
perhaps it is uniquely luminous, but additional data are needed to
establish its X-ray properties with confidence.

Monte Carlo simulations \citep{Feigelson02,Gagne04} have shown that the
formal errors on spectral fit parameters generated by {\it XSPEC} for
low-significance (faint) sources do not adequately convey the
uncertainty in these parameters.  Nevertheless, such fits are useful
for characterizing the X-ray luminosities of faint sources even though
the fit parameters are not well-known; these fits simply spline the
data and give an acceptable estimate of model normalization.  We thus
caution the reader that fit parameters for low-significance sources
should only be used to reproduce our X-ray luminosities. 

Table~\ref{tbl:apec-spec} gives thermal plasma model fits to the
stellar sources and others for which that model gave a satisfactory
fit; it is based on Tables 6--8 of \citet{Getman05}, which provides
details on how certain columns were derived.  Columns 1--4 give the
running ACIS sequence number, source name, net full-band counts, and
photometric significance from Tables~\ref{tbl:sourcelist} and
\ref{tbl:tentative}.  Columns 5--7 give the spectral fit parameters and
their 90\% confidence intervals for the absorbed thermal plasma model.
Missing confidence intervals mean that {\it XSPEC} detected some
abnormality in the error calculation from at least one of the fit
parameters; fits without uncertainties should be treated with caution.
Columns 8--12 give apparent and absorption-corrected luminosities in
various wavebands.  Column 13 completes the table with notes about the
nature of the source or the fit.

Table~\ref{tbl:powerlaw-spec} gives power law fits to those sources
outside of R136 for which a thermal plasma model gave no acceptable fit
or gave a plasma temperature that was implausibly high.  It has the
same overall structure as Table~\ref{tbl:apec-spec} but of course the
spectral fit parameters are those relevant to the absorbed power law
model and we report apparent and absorption-corrected fluxes rather
than luminosities, since the distances of many of these sources (at
least the ones that are background AGN) are unknown.

\subsection{Counterparts}

We searched for counterparts to our ACIS sources, using SIMBAD and
VizieR for sources outside R136 (off-axis angle $\theta >
30\arcsec$).   We used a progressively larger search radius $R$, with
$R=2\arcsec$ for sources with $0\arcmin.5 < \theta < 2\arcmin$,
$R=5\arcsec$ for $2\arcmin < \theta < 5\arcmin$, and $R=10\arcsec$ for
$\theta > 5\arcmin$.    When matching ACIS sources near the field
center with other catalogs of this region, counterpart searches started
by registering the ACIS astrometric reference frame and that of the
catalog under study to remove systematic offsets.  For the confused
region close to R136, we required potential counterparts to fall within
$0\arcsec.5$ of the ACIS source position in order to be considered a
match.

Table~\ref{tbl:counterparts} lists counterparts to ACIS sources from
several catalogs.  Columns 1 and 2 give the usual ACIS sequence number
and source name.  Column 3 gives the separation between the ACIS source
and its counterpart using a hierarchy of matched catalogs:  if a match
from \citet{MH94} (hereafter MH94) was available then that separation
is noted; if there was no MH94 match then the 2MASS separation is
noted; if neither of these matches exists, the separation between the
ACIS source and its match from SIMBAD or VizieR is given.  Column 4
gives the spectral type of the star or multiple stellar system if
known, taken from \citet{BAT99}, \citet{CD98} (hereafter CD98), or
SIMBAD.  Column 5 gives the \citet{BAT99} catalog number (a catalog of
LMC WR stars).  Column 6 gives the \citet{Parker93} catalog number (a
ground-based UBV photometric study of 30~Dor).  Column 7 gives the MH94
catalog number (a UBV photometric study of 30~Dor performed with the
{\em Hubble Space Telescope}).  Column 8 gives the X-ray source
identification from \citet{PPL02}.  Columns 9 and 10 give the 2MASS
match coordinates in decimal degrees (J2000).  Column 11 completes the
table with matches from other catalogs, common names for well-known
sources, and other notes.

About one third of the ACIS sources have counterparts.  Several of
these are known very early-type (WR-WR or WR-O) binaries, where the
primary source of X-rays is probably collisions between the powerful
stellar winds \citep{PPL02}.  In R136, there are at least as many
individual (not known to be binary) early O stars with detectable X-ray
emission in this observation, though, contrary to the conclusions of
\citet{PPL02}.  This will be discussed below.

\section{THE MASSIVE CLUSTER R136
\label{sec:R136}}

\begin{figure}
\centering
  \includegraphics[width=1.0\textwidth]{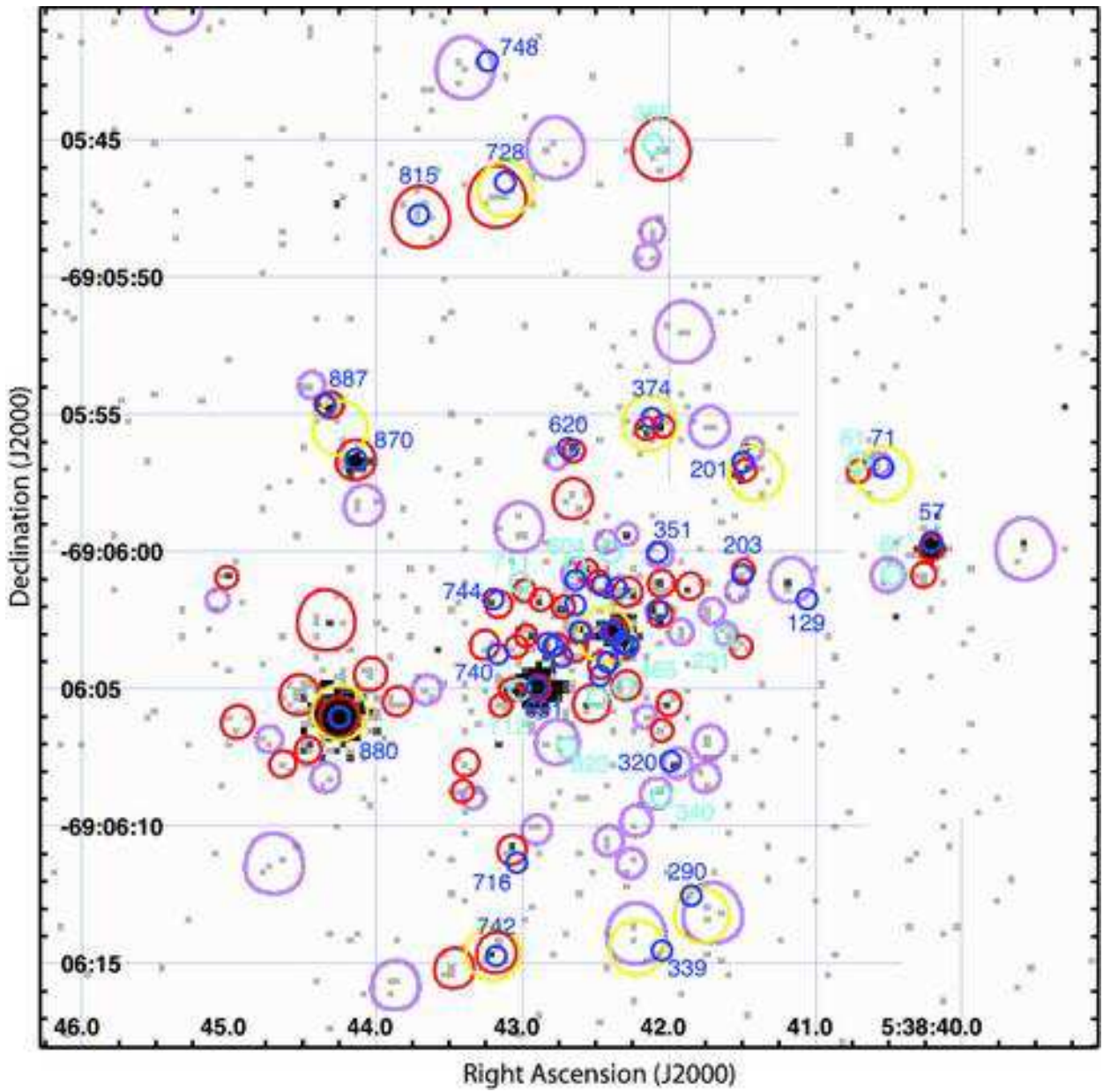}
\caption{ An image of the central part of the ACIS observation ($\sim
38\arcsec \times 38\arcsec$), binned by $0\arcsec.25$ and centered on
the massive stellar cluster R136.  Extraction polygons (as in
Figure~\ref{fig:bin4data}) for the ACIS point sources are shown in red
(primary sources) and purple (tentative sources) and are based on the
90\% PSF contour in uncrowded regions; smaller extraction polygons were
used in crowded regions to minimize cross-contamination of source
properties.  Some counterparts from 2MASS (those not suffering
confusion) are shown as large gold circles.  Counterparts from MH94 are
shown with small blue circles (brighter MH94 sources listed in CD98) and
small cyan circles (fainter sources from the MH94 catalog); their MH94
catalog numbers are given in the outer parts of the field but omitted
in the crowded central region around R136a for clarity.  All
counterparts are given in Table~\ref{tbl:counterparts}.
\label{fig:r136all}} 
\end{figure}

The X-ray point source population of 30~Dor is dominated by the central
star cluster R136.  Figure~\ref{fig:r136all} shows the environs of
R136, with ACIS source extraction regions shown in red (primary
sources) or purple (tentative sources) and some of the 2MASS
counterparts shown with large gold circles (these have been omitted in
very confused regions).  Sources from the MH94 catalog that are close
spatial matches to the ACIS sources ($<0\arcsec.5$ separation) are
indicated with small blue circles (brighter MH94 sources listed in
CD98) or small cyan circles (fainter MH94 sources).  MH94 catalog
numbers are given in unconfused regions; all MH94 numbers for matched
sources are given in Table~\ref{tbl:counterparts}.  Since we expect to
detect only the early-type members of R136 (see
\S\ref{sec:sensitivity}), only MH94 sources brighter than 16th
magnitude were considered to be possible matches to ACIS sources.  The
2MASS catalog is confused in this region; often two or more ACIS and
MH94 sources fall within the $2\arcsec$ 2MASS PSF.

\begin{figure}
\centering
  \includegraphics[width=1.0\textwidth]{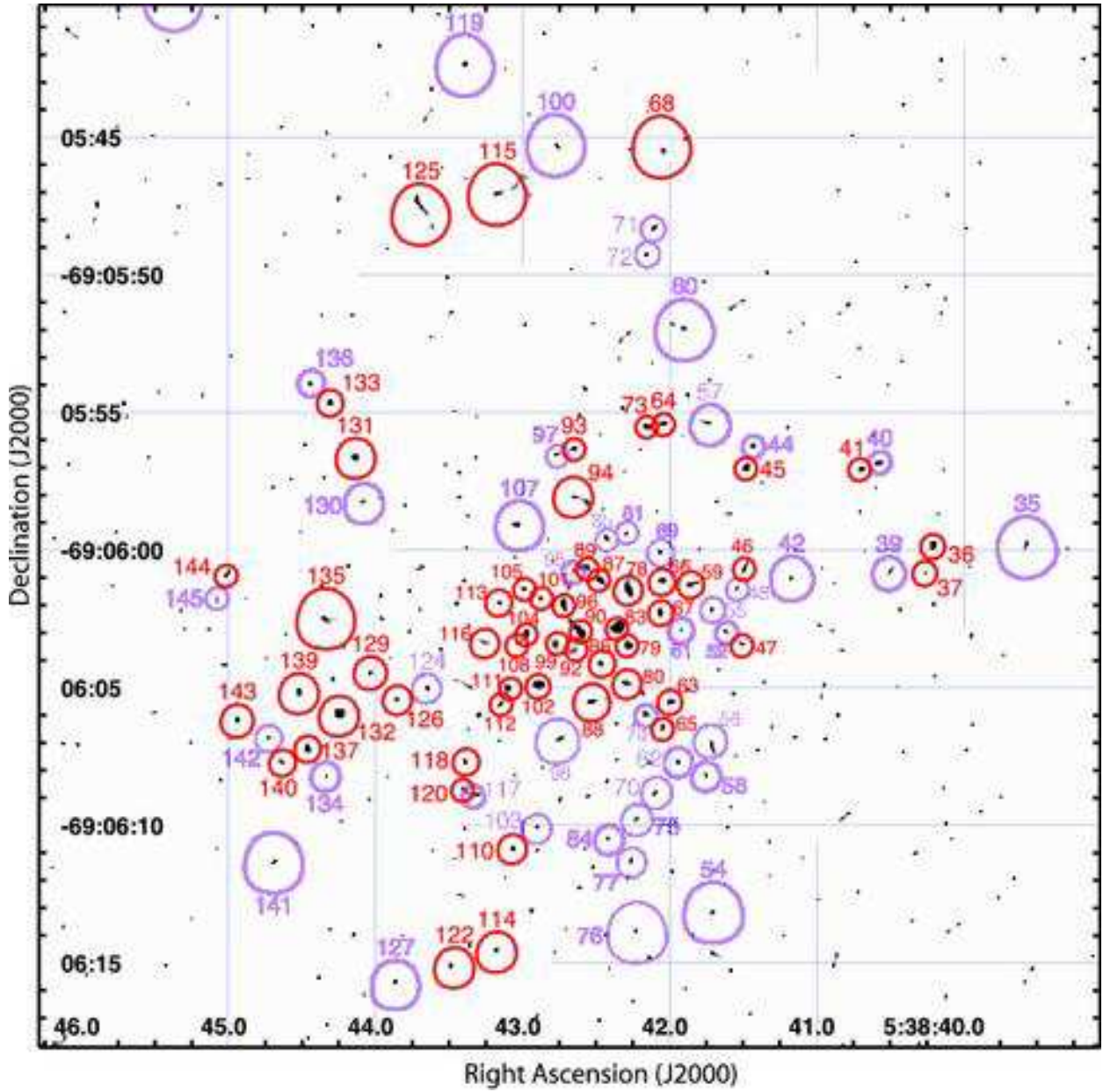}
\caption{ A maximum likelihood reconstruction (200 iterations) of the
field around R136 shown in Figure~\ref{fig:r136all}.  Extraction
polygons for the ACIS point sources that passed all validity criteria
are shown as in Figure~\ref{fig:r136all}, with primary sources in red
and tentative sources in purple.  Here counterparts have been omitted;
now ACIS sequence numbers are shown for all sources.
\label{fig:r136recon}} 
\end{figure}

One notable feature of the X-ray point source distribution
seen in Figure~\ref{fig:r136all} is the apparent clustering
of faint sources around the brightest X-ray source in the field
(CXOU~J053844.25$-$690605.9, ACIS \#132), the WR star Melnick 34 (Mk~34),
located about $10\arcsec$ east and slightly south of the R136 core and
coincident with MH94 source \#880.  None of these faint {\em Chandra}
sources has a counterpart in another waveband using our matching criteria
(although there are faint MH94 sources that are spatially coincident with
these sources) so they could be artifacts from the reconstruction (shown
in Figure~\ref{fig:r136recon}).  Since this shallow ACIS observation is
unlikely to detect lower-mass pre-main sequence stars in R136, the X-ray
companions that we see around Mk~34 are most likely individual O stars.
If these sources are confirmed in a longer X-ray observation, their
presence may indicate that Mk~34 hosts a young subcluster of massive
stars distinct from the main R136 cluster.

Figure~\ref{fig:r136recon} gives the reconstructed image of the field
shown in Figure~\ref{fig:r136all}.  From this reconstruction, we
identify 109 X-ray sources within a $30\arcsec$ radius of the core of
R136, compared to 37 sources found by a combination of {\em wavdetect}
and visual inspection of the binned image.  Figure~\ref{fig:r136recon}
shows the red (primary source) and purple (tentative source) extraction
regions for these X-ray sources along with their ACIS sequence
numbers.

Peaks in the reconstruction that lack ACIS extraction regions did not
pass our source validity criteria.  In some cases (e.g.\ ACIS \#125,
the source corresponding to MH94 \#815) the reconstruction generated
two peaks within the 90\% PSF extraction region but one of those peaks
did not pass our source validity checks, so it was included in the
extraction region of the primary peak.  A longer observation may show
two X-ray sources present in this region.

\begin{figure}
\centering
  \includegraphics[width=0.45\textwidth]{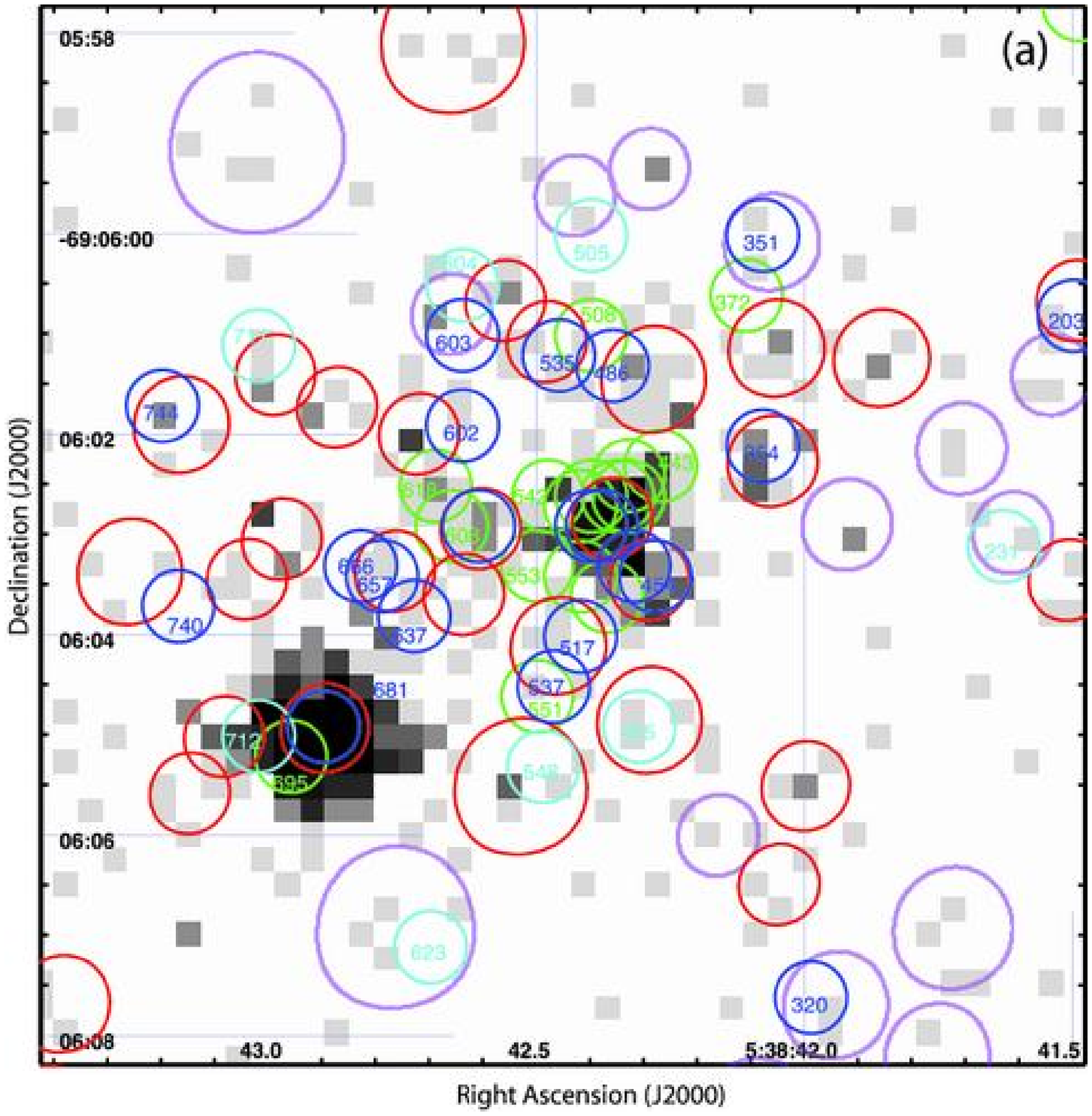}
\hspace*{0.1in}  
  \includegraphics[width=0.45\textwidth]{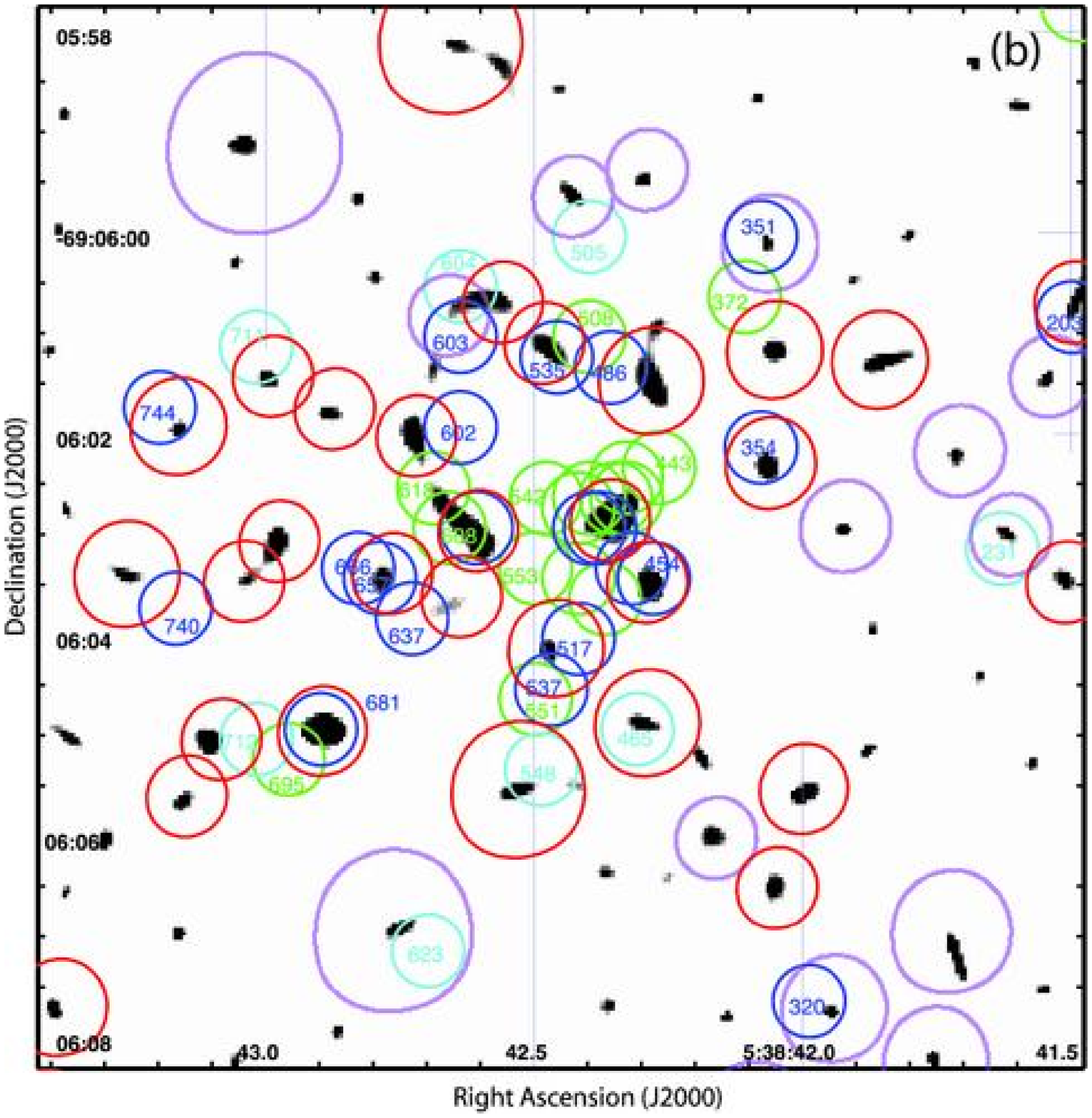}
\caption{   
(a) A zoom of Figure~\ref{fig:r136all}, showing the R136 core ($\sim
10\arcsec \times 10\arcsec$).  As in Figure~\ref{fig:r136all},
extraction polygons for ACIS sources are shown in red (primary sources)
and purple (tentative sources) and matches to the visual catalog of
MH94 are marked with circles of blue (brighter MH94 sources listed in CD98)
and cyan (fainter sources from the MH94 catalog).  Green circles now show
brighter MH94 sources (from the CD98 list) that lack {\em Chandra}
counterparts.  MH94 source names are included for all but the most crowded 
region around R136a. 
(b) The reconstructed image of the R136 core, sized to match (a) and
showing the same ACIS source regions and MH94 catalog members.
\label{fig:r136zoom}} 
\end{figure}

Figure~\ref{fig:r136zoom} shows the inner $\sim 10\arcsec \times
10\arcsec$ region around R136, displayed with $0\arcsec.25$ pixels,
before and after image reconstruction.  X-ray point sources are marked
in both images with their red or purple PSF extraction regions from
{\it ACIS Extract}.  The bright patch near field center coincides with
the dense cluster core R136a and undoubtedly represents emission from
many cluster stars.  The brightest X-ray source $4\arcsec$ to the
southeast is the WR star R136c.  Counterparts to ACIS sources  from
MH94 are shown with blue and cyan circles as in
Figure~\ref{fig:r136all}; green circles show the brightest members of
R136 in the MH94 catalog (taken from CD98) that lack individual X-ray
counterparts.

As evidenced by the swarm of green circles near R136a, in the densest
part of the cluster even image reconstruction is not sufficient to
pinpoint exactly which high-mass cluster members are X-ray emitters
above our flux limit.  There is an intriguing ring of X-ray emission
roughly $3\arcsec$ in diameter to the northeast of R136a that largely
condenses into pointlike emission in the reconstruction, although a
small ridge of emission $1\arcsec$ east of R136a remains and all the
condensations in this ring appear more extended than other point
sources in the reconstruction.  It is unclear whether this is a
physical structure or a statistical aberration due to small number
counts in these data.  Our referee suggests that it may be due to a
dense molecular cloud in this region.

Spectral properties of the brighter sources in R136 are given in
Table~\ref{tbl:apec-spec}.  They show a wide range of values, with $N_H
= 1$--$10 \times 10^{21}$~cm$^{-2}$ and $kT = 0.5$--4~keV, although the
small number of counts in most of these spectra leaves their fit
parameters not well-constrained.  These absorbing columns are typical
of massive star-forming complexes on the edges of molecular clouds.
The plasma temperatures for some sources are consistent with those seen
in single Galactic O stars \citep[$<1$~keV,][]{Chlebowski89}, although
a hotter component is sometimes present \citep{Stelzer05}.  Other R136
sources are dominated by much hotter values more typical of
colliding-wind binaries \citep{PPL02}.  This is expected given the
number of known high-mass binaries in the field \citep{BAT99}.

We also examined the average spectrum of X-ray sources in or near R136
that are individually too faint for analysis by integrating or
``stacking'' their counts.  Figure~\ref{fig:r136spec}a shows the
average spectrum of 101 sources (using 358 combined ACIS events) within
$1\arcmin$ off-axis having a photometric significance $<2$.  This
spectrum is adequately fit by a single absorbed thermal plasma with
$0.3Z_{\odot}$ abundances, $N_H \sim 0.2 \times 10^{22}$~cm$^{-2}$, and
$kT \sim 2$~keV.

This gives a coarse estimate of the average spectral characteristics of
the highest end of the X-ray luminosity function in R136.  As with the
brighter sources that we detect in this observation, the fainter
sources should be composed of early O stars, WR stars, and
colliding-wind binaries.  The spectral parameters from the stacked
spectrum are consistent with the fits to individual sources in R136
given in Table~\ref{tbl:apec-spec}, as we would expect from a similar
population.

\subsection{Melnick 34 and Other WN5h Stars
\label{sec:Mk34}}

\citet{Massey98} interpreted the WN5h sources in R136 as ``O3 stars in
Wolf-Rayet clothing,'' still burning hydrogen in their cores but so
luminous and with such prodigious mass loss that their spectra let them
masquerade as WR stars.  The brightest ACIS source in R136 is
CXOU~J053844.25$-$690605.9 (ACIS \#132), spatially coincident with Mk~34,
a 130~M$_\odot$ WN5h star located 2.62~pc from the R136 core and not
known to be binary \citep{Massey98,CD98}.  It was observed $\sim
22\arcsec$ off-axis in these ACIS data.  No variability was detected in
the X-ray lightcurve.  There are $\sim 950$ ACIS counts in the
0.5--8~keV band, or a countrate of $\sim 0.14$ counts per CCD frame.
From the simulation in \citet{Townsley02}, this countrate should not
cause substantial spectral distortion due to photon pile-up.  The
extraction region contained only 75\% of the PSF due to the presence of
several fainter sources nearby.

\begin{figure}
\centering
  \includegraphics[width=0.3\textwidth]{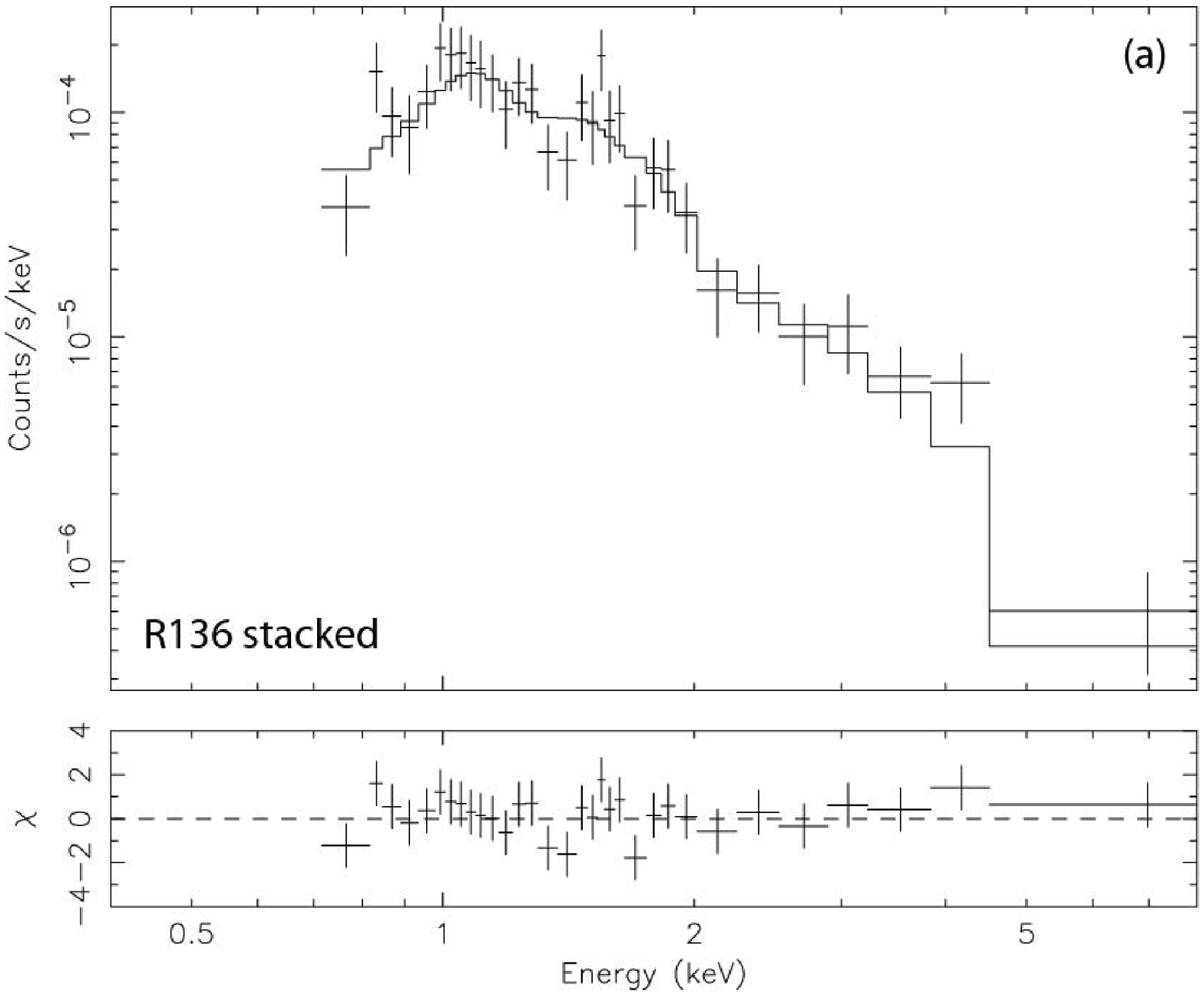}
\hspace*{0.1in}
  \includegraphics[width=0.3\textwidth]{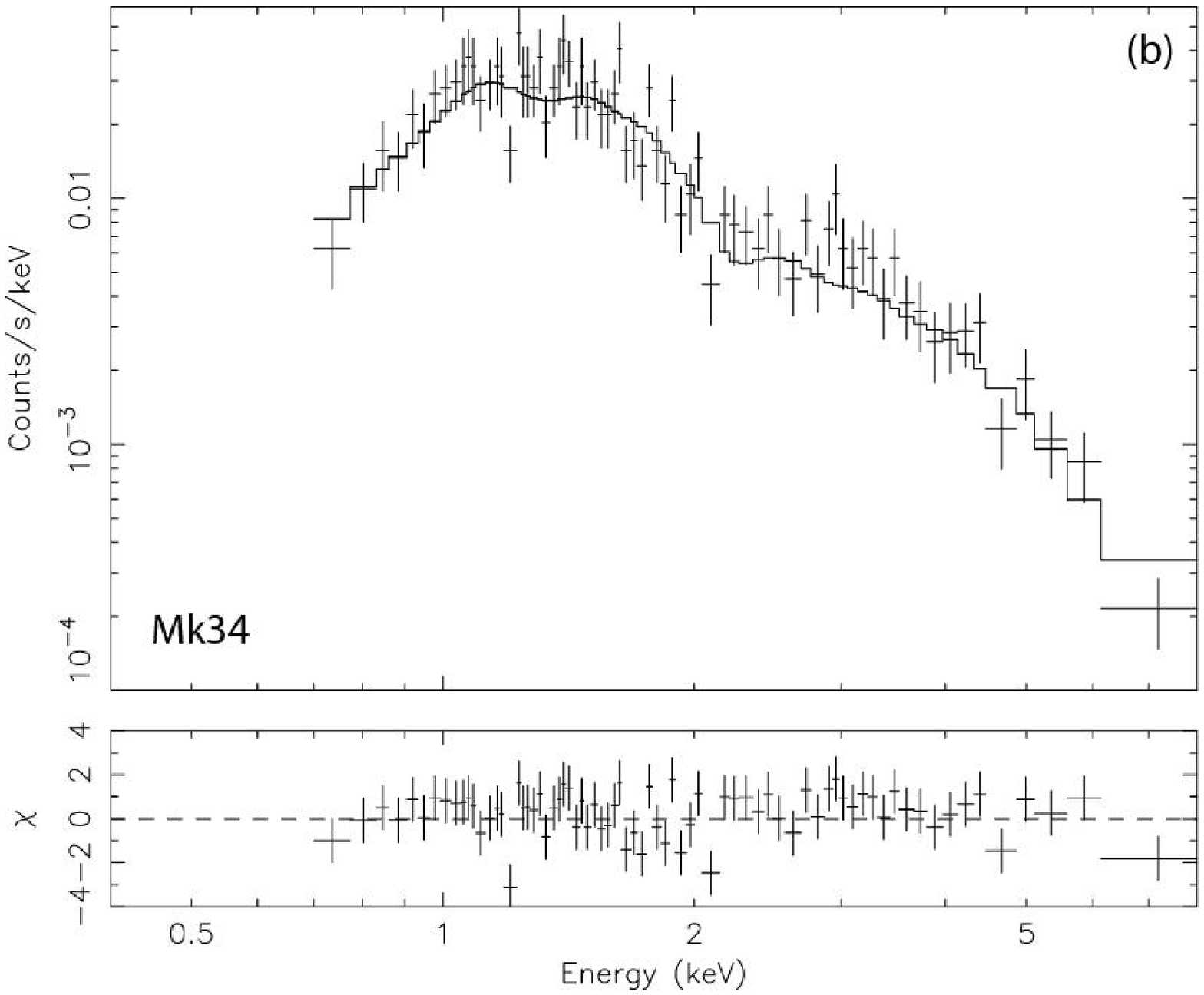}
\hspace*{0.1in}
  \includegraphics[width=0.3\textwidth]{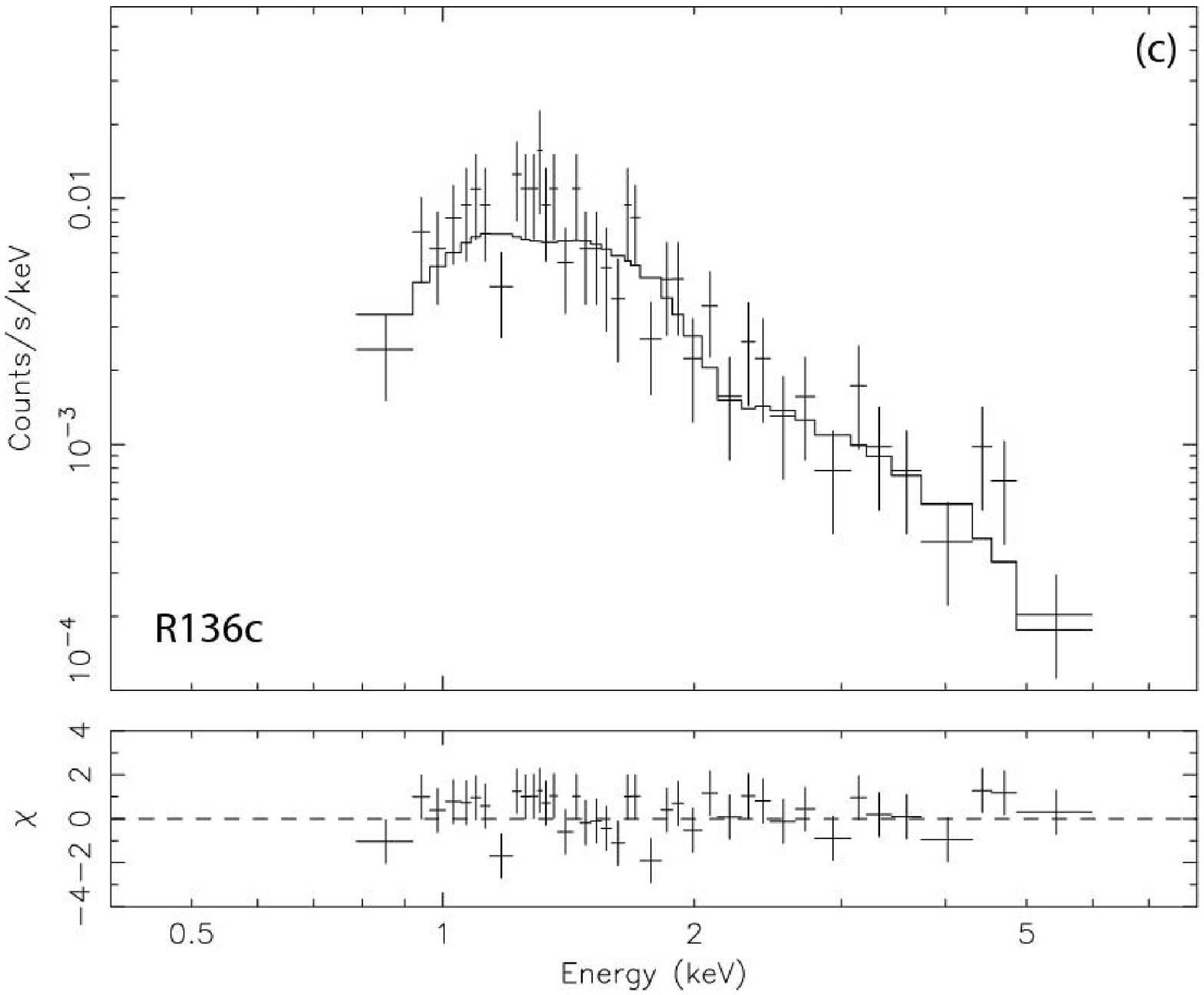}
\caption{ ACIS spectra of sources in R136 fit with absorbed thermal
plasma models.  The top panels show the data and best-fit models; the
bottom panels give the (data $-$ model) residuals.  Note that the first
panel has a different y-axis range than the other two.
(a) Average spectrum of all sources within $1\arcmin$ off-axis with
photometric significance $<2$ (generally those sources close to R136
with fewer than 10 counts).
(b) The WN5h star Mk34. 
(c) The WN5h star R136c.
\label{fig:r136spec}} 
\end{figure}

The spectrum and plasma modeling of Mk~34 derived using an {\it XSPEC}
fit to grouped data are shown in Figure~\ref{fig:r136spec}b.  This fit
is somewhat different than that from the automated spectral fitting
using ungrouped data presented in Table~\ref{tbl:apec-spec} (source
\#132) but the results are consistent to within the errors.  The model
shown in Figure~\ref{fig:r136spec}b is an absorbed {\it apec} plasma
with $N_H = 0.34^{0.43}_{0.26} \times 10^{22}$~cm$^{-2}$ and $kT =
4.3^{4.9}_{3.6}$~keV, with permitted elemental abundances ranging from
LMC to solar levels.  The absorption-corrected 0.5--8~keV luminosity of
Mk~34 is $L_{X,corr} = 2.2 \times 10^{35}$~ergs~s$^{-1}$; this agrees
well with the results of \citet{PPL02}, although our spectral fits give
a hotter, less-absorbed thermal plasma than theirs.

{\em HST} observations of Mk~34 show photometric variability of a few
tenths of a magnitude over a period of 20 days \citep{Massey02}.  Mk~34
was known to be X-ray-bright from {\em ROSAT} data \citep{Wang95}.
\citet{PPL02} find that its X-ray luminosity has not changed in the
nine years between the original {\em Einstein} observation
\citep{Wang91} and this {\em Chandra} observation.  We agree with the
conclusion of these authors, that Mk~34 is probably binary and the
source of its hard X-rays is most likely the collision of the
high-velocity winds of the binary components.

The second-brightest ACIS source in R136 is CXOU~J053842.89$-$690604.9,
with $\sim 247$ net full-band counts.  It is coincident with another
WN5h star, R136c.  It exhibits a hard spectrum similar to Mk~34
(Figure~\ref{fig:r136spec}c); the best-fit model is an absorbed {\it
apec} plasma with $N_H = 0.4^{0.6}_{0.2} \times 10^{22}$~cm$^{-2}$ and
$kT = 3.1^{5.6}_{2.0}$~keV.  The absorption-corrected 0.5--8~keV
luminosity is $L_{X,corr} = 9 \times 10^{34}$~ergs~s$^{-1}$, almost a
factor of 3 fainter than Mk~34.  As with Mk~34, we detect no X-ray
variability.

Other WN5h stars in the field are R136a1, R136a2, and R136a3
\citep{CD98}.  We detect a blend of the R136a core sources
(CXOU~J053842.35$-$690602.8) which includes components a1 and a2.  The
composite spectrum is comparatively soft ($kT = 1.3$~keV) and faint,
with just 53 net counts ($L_{X,corr} = 2 \times
10^{34}$~ergs~s$^{-1}$).  R136a3 is blended with the O3V star R136a6
and other sources yielding the composite ACIS source
CXOU~J053842.29$-$690603.4; this composite is faint (22 net counts,
$L_{X,corr} = 0.9 \times 10^{34}$~ergs~s$^{-1}$) but with a hard
spectrum ($kT = 4$~keV).

\subsection{Other R136 Stars Seen and Not Seen
\label{sec:otherstars}}

The O3If\*/WN6-A ``transition'' object Mk~39 \citep{BAT99} is detected
with 66 net counts ($L_{X,corr} = 2 \times 10^{34}$~ergs~s$^{-1}$) and
$kT = 2$~keV.  The WN6h star R134 is very faint with just 2 net counts;
due to the X-ray Eddington bias \citep{Wang04} its X-ray luminosity is
poorly constrained.  According to \citet{Massey98}, Mk~39 and R134
could be similar to the WN5h sources.  While all of these sources are
visually bright and have consistently large mass loss rates
\citep{CD98}, they range over a factor of $\sim 500$ in intrinsic X-ray
luminosity!

The O stars in R136 show similar behavior.  We detect O3 stars across a
range of luminosity classes (O3I, O3III, and O3V).  Cooler stars are
also seen:  several O5 stars, an O7V star, and the B supergiant Mk~12.
They exhibit a wide range of X-ray luminosities; while most of these
sources are quite faint, the O3III(f) star Mk~33Sa is detected with 51
net counts and fit with a soft spectrum ($kT = 0.6$~keV) and a large
absorbing column ($N_H = 1 \times 10^{22}$~cm$^{-2}$).  This
combination of spectral fit parameters leads to a large intrinsic
luminosity, $L_{X,corr} = 4 \times 10^{34}$~ergs~s$^{-1}$.  

While we detect many early-type stars in R136, there are many O3 stars
in the list of \citet{CD98} that have no X-ray counterpart at our
limiting sensitivity of $L_{X,corr} \sim 1 \times
10^{33}$~ergs~s$^{-1}$.  Notably, we do not detect 30~Dor MH36, the
O2$-$3 If\* star that helped to define the new spectral type O2
\citep{Walborn02}.  Also absent in X-rays is Mk~33Sb, the WC5 star that
\citet{Massey98} call the only {\em bona fide} WR star in their sample;
this is not surprising, since single WC stars are typically
X-ray-faint \citep{Oskinova03}.  None of the deeply embedded new
generation of massive stars \citep{Walborn02b} is detected.
  
If X-ray spectral hardness and high luminosity are indicators of close
binarity because shocks from colliding fast winds are necessary to
generate hard X-rays, we infer that Mk~34, R136c, and R136a3 are likely
close binaries, while the faint source R134 and the soft source Mk~33Sa
are not.  This contradicts somewhat the conclusions of \citet{PPL02},
who surmised that all of their {\em Chandra} sources in R136 (CX 1
through CX 9) were colliding-wind binaries.  We find that all nine of
those source locations show multiple X-ray sources within a 90\% PSF
extraction region; this is the primary source of our disagreement.

R136 provides the largest sample of such early-type stars currently
available to us.  Ideally, accurate $L_X$'s for its stars would allow
us to generate a high-mass X-ray luminosity function and extend the
$L_X$/$L_{\rm bol}$ correlation \citep{Berghoefer97} to the earliest
spectral types, but it appears that any such correlation will be masked
by the wide scatter in $L_X$ caused by details of the wind properties
and binarity in R136 stars.

\section{OTHER SOURCES IN THE FIELD
\label{sec:otherpoints}}

\subsection{Wolf-Rayet Stars}

\begin{figure}
\centering
  \includegraphics[width=0.4\textwidth]{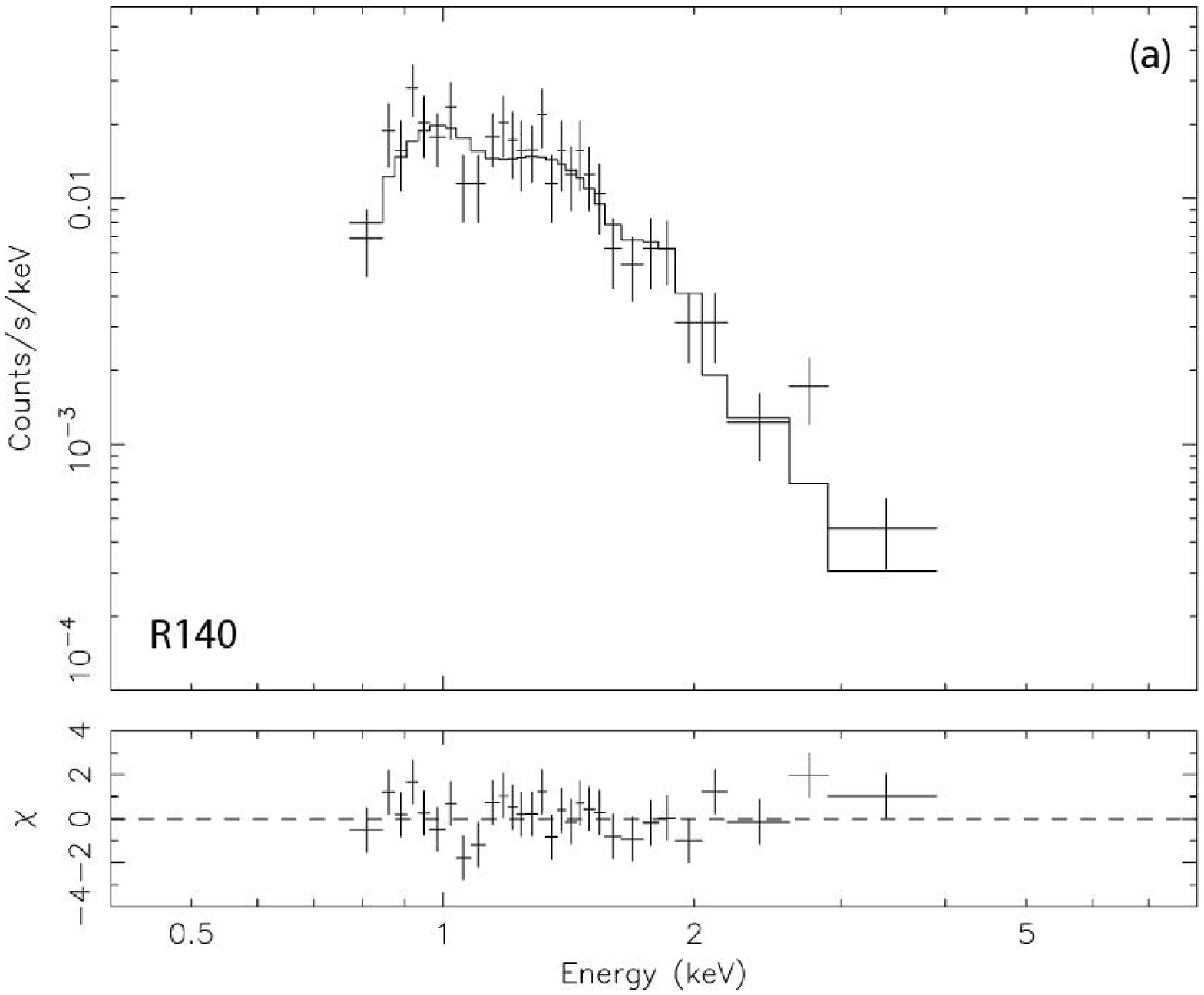}
\hspace*{0.2in}
  \includegraphics[width=0.4\textwidth]{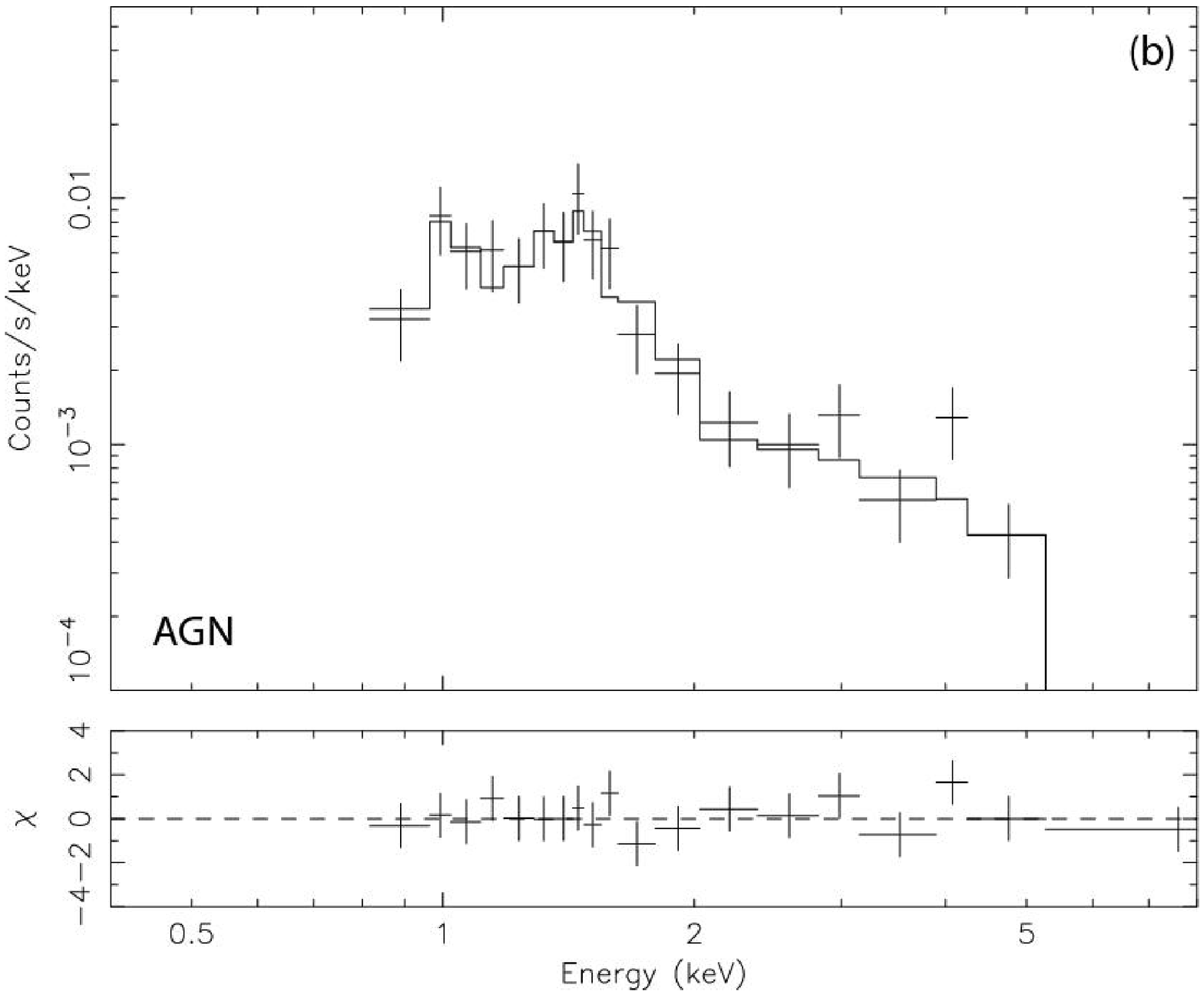}
\caption{  Spectral fits of other 30~Dor X-ray sources.  Axis ranges
match those in Figure~\ref{fig:r136spec}b and c. 
(a) The blend of massive stars R~140 a1/a2 fit by a thermal plasma
model. 
(b) A probable background AGN, CXOU~J053809.92$-$685658.3, fit by a power
law.
\label{fig:otherspec}} 
\end{figure}

We detect two sources associated with the cluster R140, imaged
$24\arcsec$ off-axis.  These were found during the first stage of
source detection; image reconstruction reveals no more.  The brighter
source (CXOU~J053841.59$-$690513.4) is coincident with the grouping of
high-mass stars R140a1/a2 \citep{Moffat87}.  It has 378 net full-band
counts ($L_{X,corr} = 1.7 \times 10^{35}$~ergs~s$^{-1}$), making it the
second brightest X-ray source in the field outside of N157B.  It is
well-fit by a thermal plasma with $kT = 0.9$~keV (see
Figure~\ref{fig:otherspec}a).  This is surprisingly soft, given the
combined spectral types of the system:  WC5+O(?) for R140a1 and WN6+O
for R140a2 \citep{BAT99}.  The WN6 star R140b is much fainter, with
just 15 net counts, but it has a much hotter (although not
well-constrained) spectrum with $kT \sim 2.5$~keV.

We also find X-ray emission associated with several other WR stars or
stellar systems spread around the field (Figure~\ref{fig:bin4data}):
R130 (WN6+B1Ia), R139 (O6Iaf/WN), R144 (WN6h), and R145 (WN6(h))
\citep[spectral types from][]{BAT99}.  All of these sources are
detected with $<20$ net counts ($L_{X,corr} \sim$2--8$\times
10^{33}$~ergs~s$^{-1}$) and $kT \sim 1.6$--2.5~keV.  Twelve of the 32
WR stars in the field from the catalog of \citet{BAT99}, most of
them single WR stars, remain undetected in this observation.

\subsection{Other Clusters}

Several X-ray sources, some with 2MASS matches, are found near the
massive stellar cluster LH~99 associated with N157B.
CXOU~J053742.64$-$690958.2 is a 54-count ACIS source that falls at the
edge of N157B's cometary tail; although it has no 2MASS match, it
coincides with UV source \#30689 from \citet{Parker98}.  Although some
of the ACIS sources are clearly associated with the SNR, some may be
massive members of LH~99.

Absent from the ACIS sourcelist are any members of the $\sim 20$~Myr
old cluster Hodge~301 \citep{Grebel00}, located $3\arcmin$ northwest of
R136.  The lack of X-ray sources in this cluster is consistent with its
age; \citet{Grebel00} note that the most massive remaining members are
early B stars and that the $\sim 40$ more massive original cluster
members may have exploded as supernovae.  Given our shallow limiting
X-ray sensitivity and the absence of O stars in this cluster, we would
not expect to detect stars in Hodge~301.

\subsection{Field 2MASS Sources}

There are several X-ray sources with 2MASS matches scattered across the
ACIS-I field that are not associated with known stars or clusters.
Many of these matches are listed as tentative in
Table~\ref{tbl:counterparts} due to separations between the ACIS source
and its proposed 2MASS counterpart that exceed $3\arcsec$.  These are
still possible matches because the ACIS sources are at large off-axis
angles (hence have large PSFs) and they are faint, so their positions
could be more uncertain than the simple error calculation in
Table~\ref{tbl:sourcelist} implies.  Nevertheless, we will not
speculate about the nature of these sources due to the uncertainty in
the matching.

Three ACIS sources (\#27, \#43, and \#171) with close 2MASS matches are
also close to UCAC2 or USNO-B sources with proper motions greater than
$\pm10$~mas~yr$^{-1}$, implying that they could be foreground stars
\citep{Momany05}.  Only \#27 is bright enough in X-rays to merit a
spectral fit; we find that it is best fit by a power law with $\Gamma =
1.7$ and high absorption, but these values are highly uncertain and the
source has only 24.1 net full-band counts.

Another four ACIS sources (\#8, \#16, \#23, and \#32) have visual as
well as 2MASS counterparts and exhibit small or unmeasured proper
motions.  They are all too faint to merit spectral fits, but \#8 has an
unusually large fraction of hard counts (see
Table~\ref{tbl:sourcelist}).  They are more likely to be LMC stars than
background sources because their $J-K$ colors are more typical of {\em
ROSAT}-selected cataclysmic variables (white dwarf X-ray binaries) and
white dwarfs than background AGN \citep[see Figure~1
of][]{Gaensicke05}.
 
If we assume that all seven described in the last two paragraphs are
foreground Galactic disk stars that serendipitously lie along the line
of sight and are coronal X-ray emitters \citep{Gudel04}, we can use
their visual and IR photometric properties to speculate on their
spectral types.  Sources \#32 and \#171 have colors consistent with
mid-M stars.  Assuming that they lie on the main sequence, their
distances are $\sim 500$~pc and $\sim 250$~pc respectively with X-ray
luminosities $L_t \sim 10^{28.5}$~ergs~s$^{-1}$, near the upper end of
the normal stellar X-ray luminosity function \citep{Schmitt95}.
Considering that they lie $3\arcsec$--$4\arcsec$ from their 2MASS
counterparts, it is possible that one or both may be spurious
identifications.

Sources \#16, \#23, and \#27 have photometric colors consistent with
mid-A stars.  If they lie on the main sequence, their distances would
be $\sim 100$, $\sim 40$, and $\sim 35$~pc, respectively.  Their
presence in the field is also difficult to understand, as the Galactic
disk population of A stars in this direction is sufficiently sparse
that not even a single such star should be present in the field,
irrespective of X-ray properties.  We therefore suggest that they may
be LMC members, perhaps A supergiants.  ACIS sources \#8 and \#43 have
extraordinary visual and IR colors inconsistent with any normal star.
We suggest that these stars are highly variable and/or exhibit strong
emission lines.  They might be cataclysmic variables in the LMC.
Visual spectroscopy of these five sources is warranted.

\subsection{X-ray Binaries}

\citet{Shtykovskiy05} surveyed several square degrees of the LMC using
{\em XMM-Newton} and found few high-mass X-ray binary systems in the
luminosity range $10^{33.5} < L_X < 10^{35.5}$~ergs~s$^{-1}$.  They
expect even fewer low-mass X-ray binaries.  Scaling their results to
our small {\it Chandra} field, we expect no bright ($L_X >
10^{35}$~ergs~s$^{-1}$) X-ray binaries in our field, and indeed find
none.  We can also exclude any high-mass X-ray binaries because no ACIS
sources outside of R136 have bright visual counterparts
\citep[see][]{Haberl02}.  Some of our sources, especially those with
very flat power law spectra ($\Gamma < 1$), could be cataclysmic
variables, such as the intermediate polars seen in profusion in the
Galactic Center \citep{Muno04}.

\subsection{Extragalactic Background Sources}

The logN-logS distribution of cosmic X-ray background sources from
\citet{Moretti03} shows that over 100 sources in our field could be from
this background population, if we were not looking through the plane
of the LMC.  Due to this additional absorption and the elevated X-ray
background due to diffuse emission in 30~Dor, we expect to see fewer
extragalactic sources.  These sources should be spread uniformly
around the field and tend to have hard spectra that are well-fit by a
power law model with photon index $\Gamma \sim 1.8$ \citep{Haberl01},
usually consistent (to within our large errors) with our sources.  Thus
we conclude that most of the ACIS sources not associated with N157B or
R136 and lacking counterparts in other wavebands are likely to be
background AGN.  There are 41 such X-ray sources in the field, plus an
additional 8 sources with only tentative 2MASS or \citet{Parker98}
matches that could be spurious.

A few of these sources can be further characterized. The bright source
CXOU~J053809.92$-$685658.3 near the northern edge of the ACIS field, also
seen by {\em ROSAT}, has a spectrum that is well-fit by a power law
with $\Gamma=1.8$ and absorption-corrected 0.5--8~keV flux $F_{t,c} =
1.5 \times 10^{-13}$~ergs~s$^{-1}$~cm$^{-2}$
(Figure~\ref{fig:otherspec}b).  This is similar to other AGN nearby
seen by {\em XMM} \citep{Haberl01}.  Our source
CXOU~J0503707.64$-$691243.4 matches source \#4 in \citet{Haberl01}, which
they also determine to be a background AGN.  One {\em Chandra} source
(CXOU~J054001.87$-$690618.7) has a counterpart in a radio catalog
\citep{Marx97} and is thus likely to be an AGN, although we find a
thermal plasma spectral fit for this source to be preferable to a power
law fit.

\section{CONCLUDING COMMENTS}

We have analyzed early {\em Chandra}/ACIS observations of 30~Doradus
and catalog here a wide variety of X-ray-emitting point sources:
individual high-mass stars, colliding-wind binaries, possibly evolved
X-ray binaries, foreground stars, and background AGN.  Using
publicly-available tools, we describe both the spatial and spectral
characteristics of these X-ray sources. 

In this $\sim 21$~ks observation of 30~Dor, 180 X-ray point sources are
found in the $17\arcmin \times 17\arcmin$ {\em Chandra}/ACIS-I field.
Most (85\%) of these sources are associated with high-mass stellar
systems.  While many of the remaining sources are likely background
AGN, a handful of white dwarf X-ray binaries may be present and one
known foreground star is seen.  Nearly 100 of the sources lie in the
R136 massive star cluster; several of these may form a small subcluster
around the WR star Mk~34.  A few sources are collected in the LH~99
cluster around the supernova remnant N157B.

Individual O stars emit X-rays at levels typical of their Galactic
counterparts; we detect a number of early-type (O3--O7) stars by using
image reconstruction to reduce the confusion in the R136 core.  The
X-ray luminosity of R136, however, is dominated by a few sources known
to be WR-WR or WR-O star binaries.  Their spectra show higher plasma
temperatures than are typical for single early-type stars; as first
noted by \citet{PPL02}, these are most likely colliding-wind binaries.
None of them show variable lightcurves in this short observation.  The
individual systems display a wide variety of X-ray luminosities,
ranging from Mk~34 with absorption-corrected $L_X = 2.2 \times
10^{35}$~ergs~s$^{-1}$ to many with $L_X < 4 \times
10^{32}$~ergs~s$^{-1}$ (assuming a typical stellar spectral model).

Thus we find no evidence for a characteristic $L_X/L_{bol}$ ratio for
the massive members of R136, such as the widely reported relation
$L_X/L_{bol} \simeq 10^{-7}$ \citep[e.g.][]{Harnden79, Pallavicini81,
Berghoefer97} that holds for later-type stars.  Rather, the X-ray
emission seems to depend critically on the details of the wind
properties and binarity for high-luminosity WR and O stars.  For
example, we infer that Mk~34, R136c, and R136a3 are likely close
binaries due to their high X-ray luminosities and/or spectral hardness,
while R134 and Mk~33Sa probably are not. 

Although our {\em Chandra} exposure is too short to penetrate deeply
into the stellar X-ray luminosity function of the region, this is still
the richest field of very-high-mass stellar sources ever characterized
in the X-ray band.  The high stellar density in R136 makes this field a
showcase for {\em Chandra's} excellent spatial resolution.  This study
marks the beginning of our attempts to understand the high-energy
processes at work in the massive stars of 30~Dor; partnered with Paper
I, we are beginning to see how the life cycle of high-mass stars shapes
the overall view of giant H{\sc II} regions and their surroundings
through their powerful winds and supernovae.  These results from early
{\em Chandra} observations lay the groundwork for more sensitive {\em
Chandra} studies planned for the near future.

\acknowledgments

Support for this work was provided to Gordon Garmire, the ACIS
Principal Investigator, by the National Aeronautics and Space
Administration (NASA) through NASA Contract NAS8$-$38252 and {\em
Chandra} Contract SV4$-$74018 issued by the {\em Chandra X-ray
Observatory} Center, which is operated by the Smithsonian Astrophysical
Observatory for and on behalf of NASA under Contract NAS8$-$03060.  We
are grateful to our referee for a thorough review of this paper and
many helpful suggestions.  LKT appreciates helpful conversations with
Mike Eracleous regarding binary populations.  This publication makes
use of data products from the Two Micron All Sky Survey, which is a
joint project of the University of Massachusetts and the Infrared
Processing and Analysis Center/California Institute of Technology,
funded by NASA and the National Science Foundation.  This research made
use of the SIMBAD database and VizieR catalogue access tool, operated
at CDS, Strasbourg, France.  We would have been lost without the
invaluable tools of NASA's Astrophysics Data System.


\onecolumn
\clearpage

\newpage

\newpage

\newpage

\newpage

\newpage

\begin{deluxetable}{rcrrrrrrrrrrrrrccccc}
\centering \rotate 
\tabletypesize{\tiny} \tablewidth{0pt}
\tablecolumns{20}

\tablecaption{ Main {\it Chandra} Catalog:  Basic Source Properties \label{tbl:sourcelist}}

\tablehead{
\multicolumn{2}{c}{Source} &
  &
\multicolumn{4}{c}{Position} &
  &
\multicolumn{5}{c}{Extracted Counts\tablenotemark{a}} &
  &
\multicolumn{6}{c}{Characteristics} \\
\cline{1-2} \cline{4-7} \cline{9-13} \cline{15-20}

\colhead{Seq} & \colhead{CXOU J} &
  & 
\colhead{$\alpha_{\rm J2000}$} & \colhead{$\delta_{\rm J2000}$} & \colhead{Err} & \colhead{$\theta$\tablenotemark{b}} &
  & 
\colhead{Net} & \colhead{$\Delta$Net} & \colhead{Bkgd} & \colhead{Net} & \colhead{PSF} &   
  &
\colhead{Signif} & \colhead{$\log P_B$\tablenotemark{c}} & \colhead{Anom\tablenotemark{d}} & \colhead{Var\tablenotemark{e}} &\colhead{EffExp} & \colhead{Med E}  \\

\colhead{\#} & \colhead{} &
  & 
\colhead{(deg)} & \colhead{(deg)} & \colhead{(\arcsec)} & \colhead{(\arcmin)} &
  & 
\colhead{Full} & \colhead{Full} & \colhead{Full} & \colhead{Hard} & 
\colhead{Frac} &
  &
\colhead{} & \colhead{} & \colhead{} & \colhead{} & \colhead{(ks)} & \colhead{(keV)}
 \\

\colhead{(1)} & \colhead{(2)} &
  & 
\colhead{(3)} & \colhead{(4)} & \colhead{(5)} & \colhead{(6)} &
  & 
\colhead{(7)} & \colhead{(8)} & \colhead{(9)} & \colhead{(10)} & \colhead{(11)} &
  & 
\colhead{(12)} & \colhead{(13)} & \colhead{(14)} & \colhead{(15)} & \colhead{(16)} & \colhead{(17)}  }

\startdata
1 & 053707.64$-$691243.4 &  &   84.281870 & $-$69.212077 &  1.2 & 11.1 &  &    20.3 &   5.9 &   8.7 &    12.1 & 0.76 &  &   3.1 & $<-$5 & ge.. & \nodata &   12.1 & 2.2 \\
2 & 053728.39$-$691217.0 &  &   84.368330 & $-$69.204740 &  1.1 &  9.4 &  &    11.2 &   5.1 &   9.8 &     6.7 & 0.91 &  &   2.0 & $-$2.9 & .... & b &   18.8 & 2.2 \\
3 & 053728.90$-$690457.9 &  &   84.370428 & $-$69.082760 &  0.7 &  6.8 &  &    11.2 &   4.4 &   3.8 &     4.0 & 0.90 &  &   2.3 & $-$5.0 & g... & \nodata &   12.0 & 1.4 \\
4 & 053736.30$-$690236.8 &  &   84.401267 & $-$69.043579 &  0.4 &  6.9 &  &    54.7 &   8.2 &   4.3 &    28.7 & 0.89 &  &   6.3 & $<-$5 & .... & a &   19.8 & 2.2 \\
5 & 053742.64$-$690958.2 &  &   84.427678 & $-$69.166169 &  0.3 &  7.0 &  &    53.8 &  14.6 & 133.2 &     0.0 & 0.89 &  &   3.6 & $<-$5 & .... & a &   19.8 & 1.2 \\
6 & 053743.45$-$690901.8 &  &   84.431083 & $-$69.150519 &  0.7 &  6.4 &  &    11.8 &   4.9 &   7.2 &     0.0 & 0.90 &  &   2.2 & $-$3.7 & .... & a &   19.9 & 1.5 \\
7 & 053745.61$-$691011.1 &  &   84.440045 & $-$69.169776 &  0.1 &  6.9 &  &  1574.3 &  49.4 & 755.7 &   535.1 & 0.89 &  &  31.5 & $<-$5 & .... & a &   19.8 & 1.6 \\
8 & 053746.01$-$690410.6 &  &   84.441717 & $-$69.069622 &  0.6 &  5.5 &  &     9.1 &   4.3 &   4.9 &     5.2 & 0.91 &  &   1.9 & $-$3.3 & .... & a &   20.2 & 2.3 \\
\enddata

\tablenotetext{a}{ Full band = 0.5--8~keV; hard band = 2--8 keV.}
\tablenotetext{b}{ Off-axis angle.}
\tablenotetext{c}{ $P_B =$ probability that extracted counts (full band) are solely from background.}
\tablenotetext{d}{ Source anomalies:  g = fractional time that source
was on a detector (FRACEXPO) is $<0.9$; e = source on field edge; p =
source piled up; s = source on readout streak.}
\tablenotetext{e}{ Source variability:  a = no evidence for
variability; b = possibly variable; c = definitely variable.  No test
is performed for sources with fewer than 4 total full-band counts.  No
value is reported for sources in chip gaps or on field edges.}

\tablecomments{See \S\ref{sec:ptsrcs} for a detailed description of
the columns.  Note that a reduced PSF fraction (significantly below 90\%) indicates that the source is in a crowded region. }

\end{deluxetable}

\newpage

\begin{deluxetable}{rcrrrrrrrrrrrrrccccc}
\centering \rotate 
\tabletypesize{\tiny} \tablewidth{0pt}
\tablecolumns{20}

\tablecaption{ Tentative Sources:  Basic Properties \label{tbl:tentative}}

\tablehead{
\multicolumn{2}{c}{Source} &
  &
\multicolumn{4}{c}{Position} &
  &
\multicolumn{5}{c}{Extracted Counts\tablenotemark{a}} &
  &
\multicolumn{6}{c}{Characteristics} \\
\cline{1-2} \cline{4-7} \cline{9-13} \cline{15-20}

\colhead{Seq} & \colhead{CXOU J} &
  & 
\colhead{$\alpha_{\rm J2000}$} & \colhead{$\delta_{\rm J2000}$} & \colhead{Err} & \colhead{$\theta$\tablenotemark{b}} &
  & 
\colhead{Net} & \colhead{$\Delta$Net} & \colhead{Bkgd} & \colhead{Net} & \colhead{PSF} &   
  &
\colhead{Signif} & \colhead{$\log P_B$\tablenotemark{c}} & \colhead{Anom\tablenotemark{d}} & \colhead{Var\tablenotemark{e}} &\colhead{EffExp} & \colhead{Med E}  \\

\colhead{\#} & \colhead{} &
  & 
\colhead{(deg)} & \colhead{(deg)} & \colhead{(\arcsec)} & \colhead{(\arcmin)} &
  & 
\colhead{Full} & \colhead{Full} & \colhead{Full} & \colhead{Hard} & 
\colhead{Frac} &
  &
\colhead{} & \colhead{} & \colhead{} & \colhead{} & \colhead{(ks)} & \colhead{(keV)}
 \\

\colhead{(1)} & \colhead{(2)} &
  & 
\colhead{(3)} & \colhead{(4)} & \colhead{(5)} & \colhead{(6)} &
  & 
\colhead{(7)} & \colhead{(8)} & \colhead{(9)} & \colhead{(10)} & \colhead{(11)} &
  & 
\colhead{(12)} & \colhead{(13)} & \colhead{(14)} & \colhead{(15)} & \colhead{(16)} & \colhead{(17)}  }

\startdata
16 & 053754.49$-$690545.5 &  &   84.477048 & $-$69.095997 &  0.5 &  4.5 &  &     5.6 &   3.4 &   2.4 &     1.3 & 0.89 &  &   1.4 & $-$2.5 & .... & a &   20.4 & 1.3 \\
20 & 053811.10$-$690523.7 &  &   84.546286 & $-$69.089943 &  0.4 &  3.0 &  &     4.4 &   3.0 &   1.6 &     0.7 & 0.90 &  &   1.2 & $-$2.2 & .... & a &   19.4 & 1.4 \\
31 & 053835.99$-$690600.3 &  &   84.649982 & $-$69.100092 &  0.4 &  0.8 &  &     3.4 &   2.5 &   0.6 &     0.0 & 0.90 &  &   1.1 & $-$2.5 & .... & a &   21.5 & 1.0 \\
32 & 053836.52$-$685726.9 &  &   84.652170 & $-$68.957477 &  1.2 &  8.3 &  &     7.4 &   4.3 &   6.6 &     1.1 & 0.90 &  &   1.5 & $-$2.1 & .... & a &   19.3 & 1.4 \\
35 & 053839.57$-$690559.8 &  &   84.664915 & $-$69.099960 &  0.2 &  0.5 &  &     3.3 &   2.5 &   0.7 &     0.9 & 0.90 &  &   1.1 & $-$2.3 & .... & a &   21.6 & 1.8 \\
38 & 053840.36$-$690532.3 &  &   84.668203 & $-$69.092306 &  0.4 &  0.4 &  &     2.6 &   2.3 &   0.4 &     0.0 & 0.90 &  &   0.9 & $-$2.0 & g... & \nodata &   13.4 & 1.0 \\
39 & 053840.51$-$690600.8 &  &   84.668795 & $-$69.100227 &  0.6 &  0.5 &  &     1.8 &   2.0 &   0.2 &     0.9 & 0.64 &  &   0.7 & $-$1.6 & .... & \nodata &   21.6 & 2.4 \\
40 & 053840.57$-$690556.7 &  &   84.669048 & $-$69.099107 &  0.2 &  0.4 &  &     1.9 &   2.0 &   0.1 &     0.0 & 0.39 &  &   0.7 & $-$2.4 & .... & \nodata &   21.6 & 1.4 \\
\enddata

\tablenotetext{a}{ Full band = 0.5--8~keV; hard band = 2--8 keV.}
\tablenotetext{b}{ Off-axis angle.}
\tablenotetext{c}{ $P_B =$ probability that extracted counts (full band) are solely from background.}
\tablenotetext{d}{ Source anomalies:  g = fractional time that source
was on a detector (FRACEXPO) is $<0.9$; e = source on field edge; p =
source piled up; s = source on readout streak.}
\tablenotetext{e}{ Source variability:  a = no evidence for
variability; b = possibly variable; c = definitely variable.  No test
is performed for sources with fewer than 4 total full-band counts.  No
value is reported for sources in chip gaps or on field edges.}

\tablecomments{See \S\ref{sec:ptsrcs} for a detailed description of
the columns.  Note that a reduced PSF fraction (significantly below 90\%) indicates that the source is in a crowded region. }

\end{deluxetable}

\clearpage


\thispagestyle{empty}

\begin{deluxetable}{rcrrrcccrcccccrc}
\centering \rotate 
\tabletypesize{\scriptsize} \tablewidth{0pt}
\tablecolumns{16}

\tablecaption{X-ray Spectroscopy for Photometrically Selected Sources:  Thermal Plasma Fits
\label{tbl:apec-spec}}

\tablehead{
\multicolumn{4}{c}{Source} &
  &
\multicolumn{3}{c}{Spectral Fit\tablenotemark{a}} &
  &
\multicolumn{5}{c}{X-ray Luminosities\tablenotemark{b}} &
  &
\colhead{Notes\tablenotemark{c}} \\ 
\cline{1-4} \cline{6-8} \cline{10-14}

\colhead{Seq} & \colhead{CXOU J} & \colhead{Net} & \colhead{Signif} &
  &
\colhead{$\log N_H$} & \colhead{$kT$} & \colhead{$\log EM$} &  
  &
\colhead{$\log L_s$} & \colhead{$\log L_h$} & \colhead{$\log L_{h,c}$} & \colhead{$\log L_t$} & \colhead{$\log L_{t,c}$} &
  &
\colhead{}  \\

\colhead{\#} & \colhead{} & \colhead{Counts} & \colhead{} &
  &
\colhead{(cm$^{-2}$)} & \colhead{(keV)} & \colhead{(cm$^{-3}$)} & 
  &
\multicolumn{5}{c}{(ergs s$^{-1}$)} &
  &
\colhead{} \\

\colhead{(1)} & \colhead{(2)} & \colhead{(3)} & \colhead{(4)} &
  &
\colhead{(5)} & \colhead{(6)} & \colhead{(7)} &
  &
\colhead{(8)} & \colhead{(9)} & \colhead{(10)} & \colhead{(11)} & \colhead{(12)} &
  &
\colhead{(13)}
}
\startdata
1 & 053707.64$-$691243.4 &    20.3 &   3.1 &  &   22.3  &  1.6  &  57.7  &  & 33.33 & 33.95 & 34.11 & 34.04 & 34.65 &  & \nodata \\
3 & 053728.90$-$690457.9 &  11.2 & 2.3 &&  \phantom{{\tiny $+0.9$}} 20.8 {\tiny $+0.9$}  & {\em  2.5 } & \phantom{{\tiny $+0.3$}} 56.3 {\tiny $+0.3$} &  & 32.99 & 32.94 & 32.94 &   33.27 & 33.32 &  & H \\
5 & 053742.64$-$690958.2 &    53.8 &   3.6 &  &   21.5  &  1.4  &  57.1  &  & 33.64 & 33.40 & 33.43 & 33.84 & 34.07 &  & \nodata \\
6 & 053743.45$-$690901.8 &    11.8 &   2.2 &  &   21.5  &  1.7  &  56.5  &  & 32.92 & 32.88 & 32.90 & 33.20 & 33.41 &  & \nodata \\
10 & 053749.11$-$690508.7 &     9.4 &   2.1 &  &   21.5  &  2.4  &  56.5  &  & 32.93 & 33.08 & 33.10 & 33.31 & 33.48 &  & R130 \\
11 & 053750.21$-$690424.0 &    10.7 &   2.1 &  &  {\em  20.7 } & {\tiny $-0.2$} 0.4 {\tiny $+0.4$} & {\tiny $-0.3$} 56.2 {\tiny $+0.4$} &  & 32.98 & 31.09 & 31.10 & 32.98 & 33.07 &  & R103f\tablenotemark{d} \\
12 & 053751.63$-$691021.4 &    99.4 &   7.4 &  &  \phantom{{\tiny $+0.6$}} 20.9 {\tiny $+0.6$} & \phantom{{\tiny $+0.1$}} 0.8 {\tiny $+0.1$} & \phantom{{\tiny $+0.3$}} 57.0 {\tiny $+0.3$} &  & 33.86 & 32.84 & 32.85 & 33.90 & 34.00 &  & \nodata \\
13 & 053752.81$-$691058.5 &    16.2 &   2.1 &  &   21.7  &  0.5  &  57.1  &  & 33.23 & 32.15 & 32.21 & 33.26 & 33.96 &  & \nodata \\
\enddata

\tablenotetext{a}{ All fits were {\it wabs*(apec)} in {\it XSPEC}, used
CSTAT, and assumed 0.3$Z_{\odot}$ abundances. }

\tablenotetext{b}{ X-ray luminosities:  s = soft band (0.5--2 keV); h =
hard band (2--8 keV); t = total band (0.5--8 keV).
Absorption-corrected luminosities are subscripted with a $c$.  When
$log N_H > 22.5$, $L_s$ and $L_{t,c}$ are unreliable and have been
omitted.  Luminosities were calculated assuming a distance of
50.0~kpc.}

\tablenotetext{c}{ H means the fit was performed by hand, usually because
the automated fit yielded non-physical results.  Well-known counterparts
from Table~\ref{tbl:counterparts} are listed here for the convenience of
the reader. }

\tablenotetext{d}{ Given luminosities are incorrect for this foreground star.
Its flux can be calculated as $\log F = \log L - 47.48$. }

\tablecomments{See \S\ref{sec:ptsrcs} for descriptions of the columns.
Quantities {\it in italics} were frozen in the spectral fit; results
are tentative.  Uncertainties represent 90\% confidence intervals.
More significant digits are used for uncertainties $<0.1$ in order to
avoid large rounding errors; for consistency, the same number of
significant digits is used for both lower and upper uncertainties.
Uncertainties are missing when {\it XSPEC} was unable to compute them
or when their values were so large that the parameter is effectively
unconstrained.  Fits lacking all uncertainties were used only as spline
fits to obtain rough estimates of luminosities; actual fit values are
unreliable. }

\end{deluxetable}

\clearpage

\thispagestyle{empty}

\begin{deluxetable}{rcrrrcccrcccccrc}
\centering \rotate 
\tabletypesize{\scriptsize} \tablewidth{0pt}
\tablecolumns{16}

\tablecaption{X-ray Spectroscopy for Photometrically Selected Sources:  Power Law Fits
\label{tbl:powerlaw-spec}}

\tablehead{
\multicolumn{4}{c}{Source} &
  &
\multicolumn{3}{c}{Spectral Fit\tablenotemark{a}} &
  &
\multicolumn{5}{c}{X-ray Fluxes\tablenotemark{b}} &
  &
\colhead{Notes\tablenotemark{c}} \\ 
\cline{1-4} \cline{6-8} \cline{10-14}

\colhead{Seq} & \colhead{CXOU J} & \colhead{Net} & \colhead{Signif} &
  &
\colhead{$\log N_H$} & \colhead{$\Gamma$} & \colhead{$\log N_{\Gamma}$} &  
  &
\colhead{$\log F_s$} & \colhead{$\log F_h$} & \colhead{$\log F_{h,c}$} & \colhead{$\log F_t$} & \colhead{$\log F_{t,c}$} &
  &
\colhead{}  \\

\colhead{\#} & \colhead{} & \colhead{Counts} & \colhead{} &
  &
\colhead{(cm$^{-2}$)} & \colhead{} & \colhead{} & 
  &
\multicolumn{5}{c}{(photons cm$^{-2}$ s$^{-1}$)} &
  &
\colhead{} \\

\colhead{(1)} & \colhead{(2)} & \colhead{(3)} & \colhead{(4)} &
  &
\colhead{(5)} & \colhead{(6)} & \colhead{(7)} & 
  &
\colhead{(8)} & \colhead{(9)} &\colhead{(10)} & \colhead{(11)} & \colhead{(12)} &
  &
\colhead{(13)}
}

\startdata
4 & 053736.30$-$690236.8 &    54.7 &   6.3 &  &  \phantom{{\tiny $+0.5$}} 21.3 {\tiny $+0.5$} & \phantom{{\tiny $+0.8$}} 1.0 {\tiny $+0.8$} & \phantom{{\tiny $+0.4$}} $-$5.3 {\tiny $+0.4$} &  & $-$14.14 & $-$13.39 & $-$13.38 & $-$13.32 & $-$13.28 &  & \nodata \\
7 & 053745.61$-$691011.1 &  1574.3 &  31.5 &  &  {\tiny $-0.06$} 21.9 {\tiny $+0.06$} & {\tiny $-0.2$} 2.4 {\tiny $+0.2$} & {\tiny $-0.08$} $-$3.3 {\tiny $+0.09$} &  & $-$12.60 & $-$12.21 & $-$12.17 & $-$12.06 & $-$11.74 &  & N157B nebula \\
9 & 053747.41$-$691019.8 &  6186.8 &  73.9 &  &  {\tiny $-0.03$} 21.8 {\tiny $+0.03$} & {\tiny $-0.07$} 2.0 {\tiny $+0.08$} & {\tiny $-0.04$} $-$2.9 {\tiny $+0.04$} &  & $-$12.04 & $-$11.54 & $-$11.51 & $-$11.42 & $-$11.21 &  & PSR~J0537$-$6910 \\
15 & 053754.39$-$690222.4 &   106.5 &   9.3 &  &  {\tiny $-0.3$} 22.0 {\tiny $+0.2$} & {\tiny $-0.6$} 2.0 {\tiny $+0.7$} & {\tiny $-0.3$} $-$4.5 {\tiny $+0.4$} &  & $-$13.88 & $-$13.21 & $-$13.16 & $-$13.12 & $-$12.85 &  & \nodata \\
19 & 053809.92$-$685658.3 &   198.8 &  12.9 &  &  {\tiny $-0.2$} 21.6 {\tiny $+0.2$} & {\tiny $-0.3$} 1.9 {\tiny $+0.5$} & {\tiny $-0.2$} $-$4.4 {\tiny $+0.2$} &  & $-$13.45 & $-$13.05 & $-$13.03 & $-$12.90 & $-$12.76 &  & \nodata \\
25 & 053826.14$-$690013.1 &    58.1 &   6.6 &  &  {\tiny $-0.6$} 21.5 {\tiny $+0.4$} & {\tiny $-0.6$} 1.8 {\tiny $+1.0$} & {\tiny $-0.3$} $-$5.1 {\tiny $+0.4$} &  & $-$14.03 & $-$13.63 & $-$13.61 & $-$13.48 & $-$13.36 &  & \nodata \\
27 & 053833.46$-$691158.7 &    24.1 &   3.6 &  &   22.1  &  1.7  &  $-$5.2  &  & $-$14.56 & $-$13.70 & $-$13.65 & $-$13.64 & $-$13.42 &  & \nodata \\
29 & 053834.79$-$685917.4 &    11.4 &   2.3 &  &   21.9  &  0.6  &  $-$6.1  &  & $-$15.10 & $-$13.86 & $-$13.84 & $-$13.84 & $-$13.78 &  & \nodata \\
\enddata

\tablenotetext{a}{ All fits were {\it wabs*(powerlaw)} in {\it XSPEC}
and used CSTAT.  $\Gamma$ is the power law photon index and
$N_{\Gamma}$ is the power law normalization.}

\tablenotetext{b}{ X-ray fluxes:  s = soft band (0.5--2 keV); h = hard
band (2--8 keV); t = total band (0.5--8 keV).  Absorption-corrected fluxes
are subscripted with a $c$.}

\tablenotetext{c}{ H means the fit was performed by hand, usually because
the automated fit yielded non-physical results.}

\tablecomments{See \S\ref{sec:ptsrcs} for descriptions of the columns.
Quantities {\it in italics} were frozen in the spectral fit; results
are tentative.  Uncertainties represent 90\% confidence intervals.
More significant digits are used for uncertainties $<0.1$ in order to
avoid large rounding errors; for consistency, the same number of
significant digits is used for both lower and upper uncertainties.
Uncertainties are missing when {\it XSPEC} was unable to compute them
or when their values were so large that the parameter is effectively
unconstrained.  Fits lacking all uncertainties were used only as spline
fits to obtain rough estimates of fluxes; actual fit values are
unreliable. }

\end{deluxetable}

\newpage

\begin{deluxetable}{rcrcccccrrrl} 
\centering \rotate
\tabletypesize{\tiny}
\tablewidth{0pt}
\tablecolumns{12}
\tablecaption{X-ray Point Source Counterparts 
\label{tbl:counterparts}}

\tablehead{
\multicolumn{2}{c}{Source} &
\colhead{Sep'n\tablenotemark{a}} &
\colhead{Spectral Type\tablenotemark{b}} &
\colhead{BAT99} &
\colhead{P93} &
\colhead{MH94} &
\colhead{PPL02} &
\multicolumn{2}{c}{2MASS Coordinates} &
  &
\colhead{Other match or other name; comments} \\ 
\cline{1-2} \cline{9-10}

\colhead{Seq \#} & \colhead{CXOU J} & 
\colhead{(\arcsec)} &
\colhead{} & 
\colhead{} &
\colhead{} &
\colhead{} &
\colhead{} &
\colhead{$\alpha_{\rm J2000}$} & \colhead{$\delta_{\rm J2000}$} &
  &
\colhead{} \\

\colhead{(1)} & \colhead{(2)} & 
\colhead{(3)} &
\colhead{(4)} & 
\colhead{(5)} &
\colhead{(6)} &
\colhead{(7)} &
\colhead{(8)} &
\colhead{(9)} & \colhead{(10)} &
  &
\colhead{(11)}   }

\startdata
 1 & 053707.64$-$691243.4 & 18.5 &   \nodata & \nodata & \nodata & \nodata & \nodata &  \nodata &  \nodata &&  {\it XMM} source \#4 from \citet{Haberl01} (ACIS source clipped)\\
 2 & 053728.39$-$691217.0 & \nodata &   \nodata & \nodata & \nodata & \nodata & \nodata &  \nodata &  \nodata &&  {\it XMM} source seen in Fig1 of \citet{Haberl01}\\
 3 & 053728.90$-$690457.9 & 4.3 &  \nodata & \nodata & \nodata & \nodata & \nodata &   84.373746 & $-$69.082619 && {\bf T}\tablenotemark{c} \\
 4 & 053736.30$-$690236.8 & 4.8 &  \nodata & \nodata & \nodata & \nodata & \nodata &   84.400018 & $-$69.042328 && {\bf T} \\
 5 & 053742.64$-$690958.2 & 3.2 &  \nodata & \nodata & \nodata & \nodata & \nodata &  \nodata &  \nodata &&  1RXH J053743.2$-$690957\tablenotemark{d}; in N157B cometary tail.\\
 6 & 053743.45$-$690901.8 & 5.4 &  \nodata & \nodata & \nodata & \nodata & \nodata &   84.434784 & $-$69.149818 && {\bf T} 1XMM J053744.7$-$690905\tablenotemark{e}\\
 7 & 053745.61$-$691011.1 & 3.2 &  \nodata & \nodata & \nodata & \nodata & \nodata &   84.438465 & $-$69.170448 && Main N157B cometary nebula\\
 8 & 053746.01$-$690410.6 & 3.3 &  \nodata & \nodata & \nodata & \nodata & \nodata &   84.439616 & $-$69.070145 && 1RXH J053745.4$-$690402\\
\enddata

\tablenotetext{a}{Source separations are with respect to (in order of availability) MH94, 2MASS, or the first entry in the ``other match'' column.}
\tablenotetext{b}{from \citet{BAT99}, \citet{CD98}, or SIMBAD}
\tablenotetext{c}{{\bf T} means match is tentative, usually due to
counterpart/source separation $> 0.5\arcsec$ for R136 sources or
separation at the edge of the 90\% {\em Chandra} PSF contour for
sources farther off-axis.}
\tablenotetext{d}{``1RXH'' sources are from the {\em ROSAT} Source Catalog of Pointed Observations with the High Resolution Imager produced by the
{\em ROSAT} Science Team in 2000.}
\tablenotetext{e}{``1XMM'' source is from the {\em XMM} Serendipitous
Source Catalog Version 1.0.1 produced by the {\em XMM-Newton} Survey
Science Centre Consortium in 2003.}
\tablenotetext{f}{``1RXS'' source is from the {\em ROSAT} Bright Source Catalog \citep{Voges99}.}
\tablenotetext{g}{From the full catalog \citet{Parker92}; too faint to appear in \citet{Parker93}.}
\tablenotetext{h}{Source confusion in the region means that other
spectral types are present in addition to those listed.}
\tablenotetext{i}{Position consistent with both 1RXH J053852.8$-$690158 and 1RXH J053853.4$-$690204.}
\tablenotetext{j}{Radio source from \citet{Marx97}.}
\tablenotetext{k}{``2RXP'' source is from the Second {\em ROSAT} (PSPC) Source Catalog of Pointed Observations, produced by the {\em ROSAT} Consortium
in 2000.}

\tablecomments{``BAT99'' = \citet{BAT99}; ``P93'' = \citet{Parker93};
``MH94'' = \citet{MH94}; ``PPL02'' = \citet{PPL02}.  Sequence numbers for tentative ACIS sources (from Table~\ref{tbl:tentative}) are shown {\em in italics}.  Counterparts have been omitted where source confusion
made it difficult to assign a one-to-one match between an ACIS source
and a counterpart.  This is particularly true in the R136 core --
\citet{Parker93} and 2MASS counterparts are generally omitted in this
region.}

\end{deluxetable}

\end{document}